\documentclass[times,review,10pt]{elsarticle}
\usepackage{orcidlink}
\usepackage{hyperref} 
\usepackage{microtype}
\usepackage{amssymb}
\usepackage{amsmath}
\usepackage{hyperref}
\usepackage{bm}
\usepackage{balance}
\usepackage[table,dvipsnames]{xcolor}
\usepackage{pifont}
\usepackage{multirow}
\definecolor{elsblue}{RGB}{10, 144, 205} 
\newcommand{\imagestyle}[2]{\setlength{\tabcolsep}{#1}\renewcommand{\arraystretch}{#2}\centering\scriptsize}
\newcommand{\tablestyle}[2]{\setlength{\tabcolsep}{#1}\renewcommand{\arraystretch}{#2}\centering}

\newcolumntype{x}[1]{>{\centering\arraybackslash}p{#1pt}}
\newcommand{\mypm}[1]{\textsubscript{\textpm #1}}
\usepackage[export]{adjustbox}

\definecolor{citecolor}{HTML}{0071BC}
\definecolor{linkcolor}{HTML}{ED1C24}
\definecolor{acceptcolor}{HTML}{74C219}
\definecolor{rejectcolor}{HTML}{DE1616}
\newcommand{\cmark}{\textcolor{acceptcolor}{\text{\ding{51}}}}
\newcommand{\xmark}{\textcolor{rejectcolor}{\text{\ding{55}}}}

\hypersetup{
            colorlinks=true,
            linkcolor=linkcolor,
            citecolor=citecolor
            }
\usepackage{caption}  
\begin{document}

\begin{frontmatter}

\title{CVP-BAA: Efficiently Adapting Segment Anything Model for Medical Image Segmentation}

\author{Mei Wang$^{\orcidlink{0009-0008-7974-6804}}$}
\ead{wm.meiwang@stu.cqut.edu.cn}
\author{Minghui Li$^{\orcidlink{0009-0004-3542-9509}}$}
\ead{minghui.li@stu.cqut.edu.cn}
\author{Libin Lan\corref{mycorrespondingauthor}$^{\orcidlink{0000-0003-4754-813X}}$}
\cortext[mycorrespondingauthor]{Corresponding author}
\ead{lanlbn@cqut.edu.cn}

\affiliation{organization={College of Computer Science and Engineering,  Chongqing University of Technology},
            city={Chongqing},
            citysep={}, 
            postcode={400054}, 
            state={Chongqing},
            country={China}}

\begin{abstract}
Various task-specific medical segmentation models have demonstrated strong performance, their heavy reliance on extensive annotations and highly specialized architectures limits their versatility, scalability, and clinical applicability. Recently, adapting vision foundation models, particularly the segment anything model (SAM), has emerged as a promising alternative. 
However, prevalent parameter-efficient fine-tuning (PEFT) strategies, such as prompt tuning and adapter tuning, often lack an explicit mechanism to embed local structural priors into the long-range dependency modeling of vision transformers (ViT), rendering them insufficiently sensitive to fine-grained details. To address these limitations, we introduce CVP-BAA, a novel hybrid PEFT method that improves SAM's sensitivity to domain-specific textures and local details, leading to superior boundary delineation. The proposed CVP-BAA adopts two key ideas. First, it introduces a convolutional visual prompting (CVP) strategy that leverages convolutional features as structural priors, prepending them to the key and value projections within the multi-head self-attention layers of strategically selected ViT blocks. Through a dedicated prompt generation-and-injection module, this mechanism effectively bridges local spatial details with global context semantics. Second, a boundary-aware adapter (BAA) is designed as a parallel branch to the feed-forward network within the same selected blocks. By explicitly computing and amplifying high-frequency components, the BAA enhances feature representations along tissue boundaries and fine-grained lesion structures. Extensive experiments on five benchmarks across three imaging modalities demonstrate that CVP-BAA achieves superior segmentation performance, surpassing existing task-specific models and PEFT-based SAM adaptations in both quantitative metrics and qualitative results. Our code is available at \href{https://github.com/WangMei-wm/CVP-BAA}{here}.
\end{abstract}


\end{frontmatter}

\section{Introduction}
\label{sec:introduction}
In recent years, deep learning has substantially advanced medical image segmentation~\cite{azad2024medical}. Convolutional models excel at capturing local spatial details, while transformer architectures are highly effective at modeling long-range dependencies~\cite{unet2015, swinunet2022}. Hybrid frameworks that integrate their respective strengths have further boosted segmentation performance~\cite{transfuse2021, lan2024brau, guo2024uctnet}. Nevertheless, most existing approaches are task-specific and rely heavily on extensive, manually curated annotations, limiting their adaptability, flexibility, and scalability in clinical practice.

To address the issue, adapting vision foundation models, particularly the segment anything model (SAM)~\cite{sam2023}, has emerged as a promising paradigm for medical image segmentation. Pretrained on massive natural image corpora, SAM has demonstrated remarkable generalization and promptable segmentation capabilities~\cite{Wu2025Med-SA}. 
However, directly applying SAM to medical imaging often yields suboptimal performance~\cite{SAMed2023, AFTer-SAM2024}. This gap primarily stems from a substantial domain shift between natural and medical images. Medical images frequently exhibit modality-specific characteristics, such as low tissue contrast, pervasive noise, and imaging artifacts, that are underrepresented in natural image pretraining distributions. Furthermore, SAM's underlying vision transformer (ViT) encoder, while highly effective at modeling global dependencies, inherently lacks explicit mechanisms to preserve local spatial structures and fine-grained boundary details~\cite{sam2023, xu2021vitae}, yet clinical segmentation demands precise delineation of subtle anatomical structures and irregular lesion boundaries. As a result, SAM often struggles to balance global semantics with fine-grained structural information in medical imaging scenarios. Therefore, developing locality-aware adaptation strategies is essential to efficiently transfer SAM for precise medical image segmentation.

Recent studies have explored various fine-tuning strategies to better adapt SAM for medical image segmentation~\cite{SAMed2023, Medsam2024, Wu2025Med-SA}. A straightforward approach is full fine-tuning --- updating all parameters of the SAM model. However, this is often impractical or infeasible for large-scale ViT architectures due to their high computational cost, which becomes increasingly prohibitive as model size grows. Moreover, full fine-tuning on limited medical datasets carries significant risks of catastrophic forgetting and overfitting~\cite{samus2024, liu2025medsam3}. To address these constraints under data-scarce conditions, MedSAM~\cite{Medsam2024} employs a partial component fine-tuning strategy~\cite{ye2023partialfinetuning} by freezing the encoder and updating only the mask decoder. This approach demonstrates improved segmentation performance over the original SAM on medical images. Another line of work leverages parameter-efficient fine-tuning (PEFT) techniques to adapt SAM with minimal trainable parameters. These include low-rank adaptation (LoRA)~\cite{Hu2022LoRA}, visual prompt tuning (VPT)~\cite{vpt2022}, and adapter modules~\cite{Houlsby2019Adapter}. Representative PEFT-based methods for medical segmentation include SAMed~\cite{SAMed2023}, which applies LoRA to the ViT encoder; SAMUS~\cite{samus2024}, which augments the encoder with bottleneck adapters and a parallel convolutional neural network (CNN) branch; and Med-SA~\cite{Wu2025Med-SA}, which employs prompt-conditioned adapters with space-depth transpose for 3D volumetric adaptation. Despite these advances, existing adaptation strategies often fail to fully capture the domain-specific characteristics of medical images, such as low contrast, heterogeneous textures, and complex anatomical structures~\cite{moglia2025generalist}, and may still yield suboptimal results, particularly when segmenting targets with intricate or ambiguous boundaries.

\begin{figure}[!t]
\centering
\includegraphics[width=1.0\textwidth]{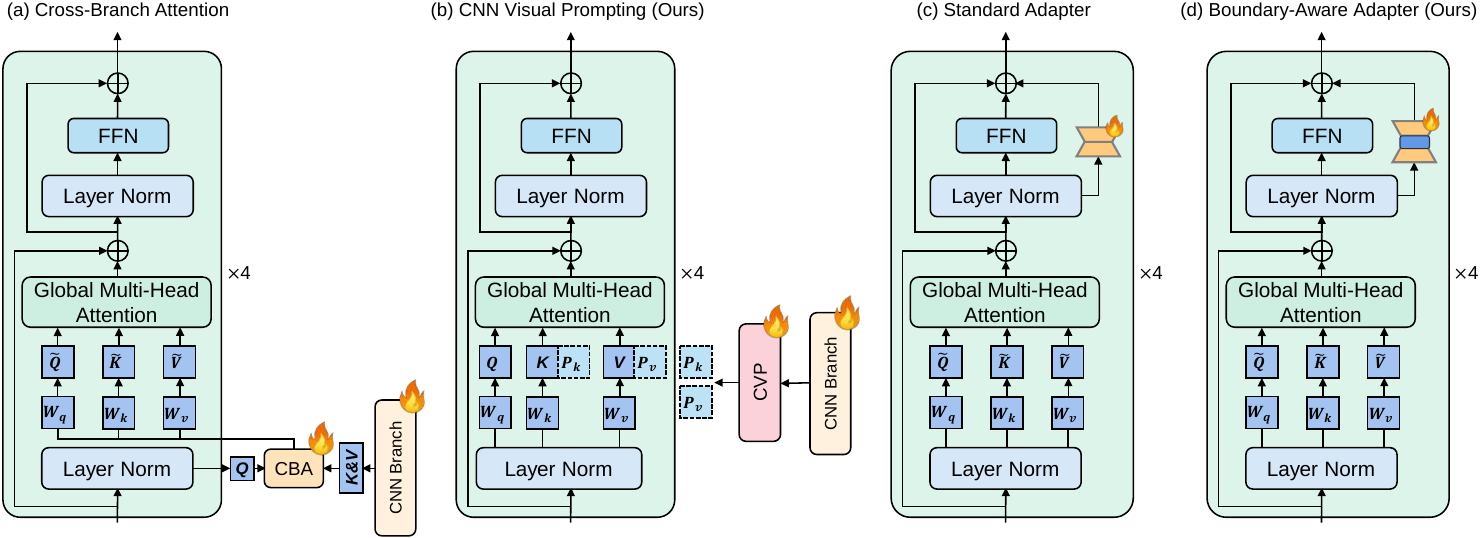}
\caption{Motivation behind the proposed CVP-BAA framework. (a) Existing methods such as SAMUS inject local CNN features into the ViT image encoder via cross-branch attention; however, this implicit fusion fails to provide explicit guidance for the global self-attention mechanism. (b) To address this, we introduce convolutional visual prompting (CVP) directly into the ViT self-attention layers, enabling local structural priors to explicitly guide global dependency modeling. (c) Conventional adapter-based PEFT methods typically insert standard adapters into the SAM ViT blocks, which inherently lack local inductive biases and explicit boundary-aware modeling. (d) In contrast, we propose replacing the standard adapter with a boundary-aware adapter (BAA), a parallel component alongside the feed-forward network (FFN). The BAA explicitly extracts and amplifies high-frequency boundary cues, thereby ensuring the precise delineation of ambiguous regions and fine-grained structures.}
\label{fig:motivation}
\end{figure}

As illustrated in Fig.~\ref{fig:motivation}, several recent studies have sought to incorporate local information, typically extracted via CNNs, into attention mechanisms through cross-attention modules~\cite{samus2024, Qin2024DBSAM}, such local cues are generally not integrated into the self-attention computation itself. Consequently, the implicit infusion of local information remains insufficient for achieving precise, fine-grained segmentation of intricate anatomical structures and lesion boundaries in medical imaging. In addition, many existing adapter-based methods usually originate from generic adaptation paradigms developed for computer vision or natural language processing tasks, and thus fail to explicitly account for the structural characteristics and texture details inherent to medical images ~\cite{Wu2025Med-SA, Hu2022LoRA, Chen2022AdaptFormer}. Motivated by these observations, we propose a natural yet effective strategy: explicitly guiding global dependency modeling with local priors while simultaneously enhancing boundary-aware local structural refinement during SAM adaptation. To this end, we introduce \textbf{CVP-BAA}, a novel hybrid PEFT method that improves SAM's sensitivity to domain-specific textures and fine-scale details, thereby enabling superior boundary delineation in medical image segmentation.

The main contributions of this work are summarized as follows:
\begin{itemize}
    \item We propose a novel \textbf{C}onvolutional \textbf{V}isual \textbf{P}rompt Tuning (\textbf{CVP}) strategy that explicitly injects local structural priors into the ViT's global attention mechanism. Specifically, CNN-derived features are transformed into visual prompts and prepended to the key and value sequences of the multi-head self-attention (MHSA) layers within strategically selected transformer blocks. Equipped with a dedicated prompt generation-and-injection module, CVP effectively augments long-range semantic modeling with fine-grained local details.
    \item We design a \textbf{B}oundary-\textbf{A}ware \textbf{A}dapter (\textbf{BAA}) as a parallel branch to the feed-forward network (FFN) within the same selected ViT blocks. BAA explicitly computes and amplifies high-frequency components such as edges and textures via a boundary enhancement (BE) module and branch scaling factor $s$ to enrich feature representations along tissue boundaries and fine-grained anatomical structures, thereby mitigating the baseline model's insensitivity to edge-level details.
    \item Extensive experiments across five benchmarks spanning three medical imaging modalities demonstrate that CVP-BAA consistently achieves state-of-the-art segmentation performance, outperforming both existing task-specific models and PEFT-based SAM adaptation methods in terms of both quantitative metrics and qualitative results.
\end{itemize}

The remainder of this paper is organized as follows. Section~\ref{sec:relatedwork} delves into previous relevant work and their limitations on medical image segmentation tasks. Section~\ref{sec:method} details our model architecture and the innovations of our method. Section~\ref{sec:experimentalsets} presents the experimental details, including quantitative and qualitative results that compare our method with task-specific segmentation  models and SAM adaptation baselines, and provides an in-depth analysis of the results. Section~\ref{sec:conclusion} summarizes our findings, points out the limitations of the work, and indicates directions for future research.

\section{Related Work}
\label{sec:relatedwork}
\subsection{Task-Specific Approaches}
\label{segmentation_model}
Medical image segmentation is a cornerstone of computer-aided diagnosis, enabling precise anatomical delineation and lesion quantification. Over the past decade, deep learning has driven substantial progress through several evolving architectural paradigms. Initially, convolutional neural networks (CNNs) dominated the field by leveraging strong inductive biases such as local receptive fields and translation equivariance. Representative U-shaped architectures like U-Net~\cite{unet2015}, CE-Net~\cite{ce-net2019}, and AAU-Net~\cite{aaunet2022} effectively preserve spatial details through skip connections and multi-scale feature aggregation. Despite these advancements, the inherent locality of convolutional operations fundamentally restricts their ability to model long-range dependencies, often leading to suboptimal performance when segmenting targets with large shape variations or complex lesion structures. 

To overcome this issue, transformer-based architectures have been increasingly adopted. Models such as Swin-Unet~\cite{swinunet2022} and MISSFormer~\cite{missformer2021} replace convolutional encoders with hierarchical self-attention mechanisms, enabling global context modeling across the entire image. Nevertheless, such pure transformers lack inherent spatial inductive biases, making them prone to over-smoothing fine-grained boundaries and struggling with low-contrast or small-scale medical targets, especially under limited annotations. Thus, hybrid CNN-Transformer architectures have emerged to combine local feature extraction with global semantic modeling. TransUNet~\cite{chen2024transunet} pioneers this direction by feeding CNN-extracted feature maps into a transformer encoder and fusing high-resolution CNN features via skip connections. TransFuse~\cite{transfuse2021} adopts a parallel dual-branch design, integrating CNN and transformer pathways through a bidirectional fusion module to simultaneously capture local details and global semantics. Similarly, H2Former~\cite{h2former2023} constructs a hierarchical hybrid block that unifies convolution, multi-scale channel attention, and token-wise self-attention within a single stage. While these hybrid designs improve segmentation accuracy, they often rely on complex, manually engineered fusion strategies and remain heavily dependent on task-specific architectural tuning.

Furthermore, the quadratic computational complexity of vanilla self-attention has spurred research into efficient attention mechanisms. Recent works like BRAU-Net++~\cite{lan2024brau}, MedFormer~\cite{Xia2025medformer}, and MSLAU-Net~\cite{lan2026mslaunet} introduce sparse~\cite{child2019generatingls, zhao2019explicit} or linear attention~\cite{katharopoulos2020transformers} to scale to higher resolutions. Although these methods significantly reduce computational burdens, they still require extensive task-specific architectural redesign and struggle to generalize across diverse imaging modalities without retraining. This motivates a move away from task-specific architectural engineering toward adapting vision foundation models.

\subsection{Segment Anything Model Adaptations}
\label{fine-tuning}
While adapting SAM for domain-specific tasks has garnered more attention, achieving effective and robust adaptation for medical image segmentation remains a critical challenge.
PEFT has emerged as a promising paradigm, freezing the majority of pretrained weights while updating only a minimal, targeted subset. This strategy drastically reduces computational and memory overhead while preserving the rich, domain-agnostic representations learned in the foundation model. Existing PEFT-based adaptations for SAM can be broadly categorized into three representative categories: prompt-based tuning~\cite{vpt2022,prompttuning2021}, adapter-based tuning~\cite{chen2023sam-adapter, pfeiffer2021adapterfusion}, and low-rank adaptation~\cite{Hu2022LoRA}.

Prompt-based methods, such as TP-SAM~\cite{ptsam2025} and BT-SAM~\cite{li2026btsam}, introduce learnable continuous vectors into the input or embedding space to steer SAM toward downstream tasks. While highly parameter-efficient and effective at preserving pretrained knowledge, these prompts primarily operate at the input level. 
Consequently, they offer limited capacity to explicitly model local texture patterns, fine-grained anatomical structures, and ambiguous boundaries that are paramount in medical imaging.
Adapter-based strategies insert lightweight bottleneck modules into transformer layers to bridge the domain gap. Early approaches adopted the conventional adapter design~\cite{Houlsby2019Adapter}, which, despite its efficiency, lacks domain-specific inductive biases and often struggles to capture intricate local details. To mitigate this, recent works such as Trans-SAM~\cite{transsam2025}, NAS-Adapter~\cite{chen2026nsa-adapter}, CC-SAM~\cite{gowda2024cc-sam} and SAMUS~\cite{samus2024} have integrated convolutional operations or parallel CNN branches into the adapter architecture, thereby injecting local spatial priors and enhancing boundary perception. Low-rank adaptation (LoRA)~\cite{Hu2022LoRA} represents another dominant PEFT paradigm, approximating weight updates via low-rank matrix decomposition. Recent works like SAMed~\cite{SAMed2023} apply LoRA to the query and value projection layers of SAM’s image encoder, achieving competitive performance with minimal trainable parameters.

Despite these advances, two fundamental limitations remain. First, existing methods predominantly integrate local cues through post-hoc convolutional enhancement or cross-branch fusion, rather than explicitly injecting local structural priors into the core self-attention mechanism. As a result, their influence on the ViT’s long-range dependency modeling remains suboptimal. To address this, we propose the convolutional visual prompting (CVP) strategy, which explicitly prepends CNN-derived local priors to the key and value sequences within multi-head self-attention. Second, most lightweight modules prioritize generic low-dimensional feature transformations, lacking explicit mechanisms to enhance high-frequency boundary information. This deficiency is particularly detrimental for medical images characterized by low contrast, blurred edges, and subtle texture transitions. Meanwhile, recent frequency-domain analyses in foundation models reveal that task-relevant semantics are usually encoded in high-frequency features, while domain-specific biases reside in low-frequency components~\cite{lee2026deep, imedsam2024}. Inspired by this, we design a boundary-aware adapter (BAA) that explicitly computes and amplifies high-frequency components to improve feature representations along lesion boundaries.

\section{Method}
\label{sec:method}
\subsection{Overall Framework}
\begin{figure}[htbp]
\centering
\includegraphics[width=1.0\linewidth]{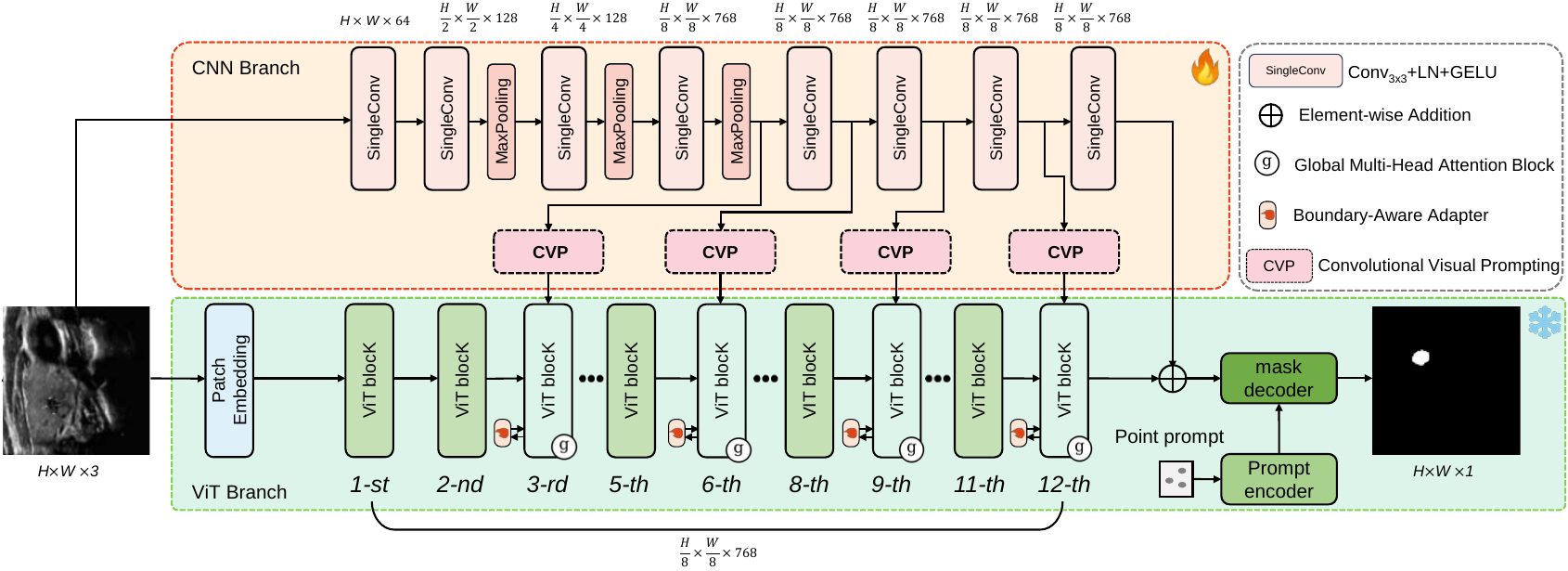}  
\caption{Architecture overview of the proposed CVP-BAA framework for efficient SAM adaptation. The framework consists of two core components: the convolutional visual prompting (CVP) module, which explicitly injects local structural priors into the global self-attention layer, and the boundary-aware adapter (BAA), which isolates and amplifies high-frequency boundary cues. By combining these components, CVP-BAA achieves parameter-efficient adaptation of the frozen SAM backbone, yielding highly precise and robust medical image segmentation.}
\label{fig:architect}  
\end{figure}

As illustrated in Fig.~\ref{fig:architect}, our framework preserves the foundational SAM architecture while systematically enhancing its image encoder to better capture local anatomical structures and boundary details critical for medical image analysis. Inspired by the dual-branch encoding paradigm of SAMUS~\cite{samus2024}, we adopt a parallel feature extraction scheme comprising a \textit{ViT branch} and a complementary \textit{CNN branch}.

Given an input image of resolution $H$$\times$$W$$\times$3, where $H$ and $W$ denote the height and width, respectively, both branches process the image in parallel.
In the\textit{ ViT branch}, a patch embedding layer projects the input into a sequence of tokens, establishing a feature resolution of ${H/8}$$\times$${W/8}$, which remains consistent across all subsequent transformer blocks. Meanwhile, the \textit{CNN branch} employs a lightweight pathway consisting of one \texttt{SingleConv} layer and three successive \texttt{SingleConv}-\texttt{MaxPooling} blocks, followed by four additional \texttt{SingleConv} layers. Each \texttt{SingleConv} block comprises a 3$\times$3 convolution, layer normalization, and a GELU activation. This design progressively downsamples the feature maps and refines their representations, ultimately aligning their spatial resolution with that of the ViT branch at ${H/8}$$\times$${W/8}$.

To address the inherent limitation of vanilla self-attention in capturing fine-grained local details, we integrate two novel modules into specific global attention transformer blocks. 
First, we introduce convolutional visual prompting (CVP), as shown in Fig.~\ref{fig:prompt}, a mechanism that explicitly extracts local structural priors from the outputs of the final \texttt{SingleConv}-\texttt{MaxPooling} block and the subsequent three \texttt{SingleConv} blocks in the CNN branch.
These priors are injected as visual prompts by being prepended to the key and value projections within the multi-head self-attention (MHSA) layers of the \textit{3}-rd, \textit{6}-th, \textit{9}-th, and \textit{12}-th ViT blocks, thereby enriching long-range dependency modeling with structural cues. Second, we propose the boundary-aware adapter (BAA), as depicted in Fig.~\ref{fig:adapter}, a parallel adaptation module inserted alongside the feed-forward network (FFN). The BAA is specifically designed to compute and amplify high-frequency components of the features, significantly enhancing the representation of tissue boundaries and other subtle anatomical structures. 
Together, CVP and BAA enable the efficient adaptation of SAM, leading to superior segmentation performance.

\subsection{Convolutional Visual Prompting}
\label{subsec:GI_prompt_block}
\begin{figure}
\centering
\includegraphics[width=1.0\linewidth]{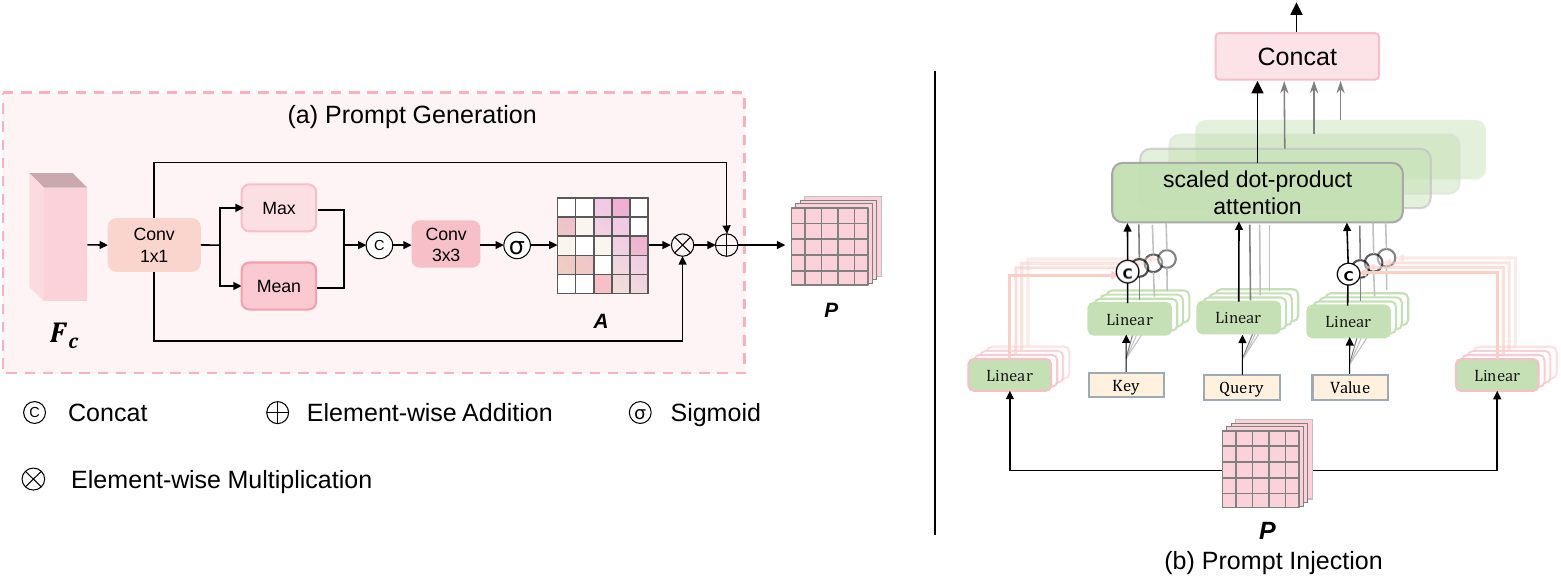}
\caption{Illustration of the proposed convolutional visual prompting (CVP) component. The CVP consists of two sequential stages: (a) prompt generation, which extracts fine-grained local structural priors via a parallel CNN branch, and (b) prompt injection, which explicitly integrates these local representations into the ViT's self-attention mechanism to guide global dependency modeling.}
\label{fig:prompt}
\end{figure}

The convolutional visual prompting (CVP) component is designed to inject local structural representations into global dependency modeling. As illustrated in Fig.~\ref{fig:prompt}, it comprises two stages: prompt generation and prompt injection.

\textbf{Prompt Generation}. As shown in Fig.~\ref{fig:prompt} (a), let $\bm{F}_c \in \mathbb{R}^{N \times C}$ denote the feature map produced by the corresponding \texttt{SingleConv} (-\texttt{MaxPooling}) block of the parallel CNN branch, where $N=HW$ is the spatial sequence length and $C$ represents the channel dimension. To enhance local structural representations, we first refine $\bm{F}_c$ using a $1 \times 1$ convolution to obtain $\hat{\bm{F}}_c$. Subsequently, a spatial attention mechanism is employed to highlight salient lesion regions. Specifically, average pooling and max pooling are performed on $\hat{\bm{F}}_c$ along the channel dimension, formulated as:
\begin{equation}
\bm{F}_{\mathrm{avg}} = \mathrm{AvgPool}(\hat{\bm{F}}_c), \quad
\bm{F}_{\mathrm{max}} = \mathrm{MaxPool}(\hat{\bm{F}}_c),
\end{equation}
where $\bm{F}_{\mathrm{avg}}, \bm{F}_{\mathrm{max}} \in \mathbb{R}^{N \times 1}$.
These two feature maps are then concatenated along the channel dimension and processed through  a 3$\times$3 convolution followed by a Sigmoid activation function to generate a spatial attention map:
\begin{equation}
\bm{A} = \sigma \left(\mathrm{Conv}_{3\times3}([\bm{F}_{\mathrm{avg}}; \bm{F}_{\mathrm{max}}]) \right),
\end{equation}
where $\bm{A} \in \mathbb{R}^{N \times 1}$ encodes the learned importance weights, and $[\cdot;\cdot]$ denotes channel-wise concatenation. Finally, the refined visual prompt $\bm{P}\in \mathbb{R}^{N \times C}$  is obtained by adaptively weighting the local features:
\begin{equation}
\bm{P} = \hat{\bm{F}}_c + \hat{\bm{F}}_c \otimes \bm{A},
\end{equation}
where $\otimes$ denotes element-wise multiplication. This design effectively amplifies boundary-sensitive local structures while preserving the original feature semantics, yielding robust prompts for subsequent injection into the global attention mechanism.

\textbf{Prompt Injection}.
As illustrated in Fig.~\ref{fig:motivation} (b) and Fig~\ref{fig:prompt} (b), unlike SAMUS~\cite{samus2024}, which injects CNN features via cross-branch attention, we explicitly construct refined visual prompts $\bm{P} \in \mathbb{R}^{N \times C}$ to generate key and value prompts, denoted as $\bm{P}_k\in \mathbb{R}^{N \times C}$ and $\bm{P}_v\in \mathbb{R}^{N \times C}$. These are directly prepended to the original key ($\bm{K}\in \mathbb{R}^{N \times C}$) and value ($\bm{V}\in \mathbb{R}^{N \times C}$) sequences within the self-attention layers.

To enable parallel multi-head computation, we first split the input features and prompts into $h$ attention heads via a linear operation. Let $\bm{Q}^i, \bm{K}^i, \bm{V}^i \in \mathbb{R}^{ N \times d_h}$ denote the query, key, and value representations for the $i$-th head ($i = 1, \ldots, h$), where $d_h=C/h$ is the dimension of each head. Similarly, the prompts are partitioned into $\bm{P}^i_k, \bm{P}^i_v \in\mathbb{R}^{ N \times d_h}$. The projections for the $i$-th head are computed as:
\begin{equation}
   \bm{Q}^i=\bm{Q}{\bm{W}^i_q}, \quad \bm{K}^i=\bm{K}{\bm{W}^i_k}, \quad \bm{V}^i=\bm{V}{\bm{W}^i_v}, 
\end{equation}
and the corresponding prompt projections are obtained by:
\begin{equation}
   \bm{P}^i_k=\bm{P}{\bm{W}^i_{pk}}, \quad \bm{P}^i_v=\bm{P}{\bm{W}^i_{pv}}, 
\end{equation}
where $\bm{W}^i_q, \bm{W}^i_k, \bm{W}^i_v, \bm{W}^i_{pk}, \bm{W}^i_{pv} \in \mathbb{R}^{C \times d_h}$ are the learnable projection matrices for the $i$-th head.
We then concatenate these prompts along the sequence dimension to augment the original sequences:
\begin{equation}
\tilde{\bm{K}}^i = [\bm{K}^i; \bm{P}^i_k], \quad \tilde{\bm{V}}^i = [\bm{V}^i; \bm{P}^i_v],
\end{equation}
where $[\cdot;\cdot]$ denotes concatenation along the sequence dimension, yielding updated key and value tensors $\tilde{\bm{K}}^i, \tilde{\bm{V}}^i \in \mathbb{R}^{2N \times d_h}$.

The attention output for each head is then computed via the standard scaled dot-product mechanism:
\begin{equation}
\mathrm{head}_i = \mathrm{Attention}(\bm{Q}^i, \tilde{\bm{K}}^i, \tilde{\bm{V}}^i) = \mathrm{Softmax}\left( \frac{\bm{Q}^i (\tilde{\bm{K}}^i)^\top}{\sqrt{d_h}} \right) \tilde{\bm{V}}^i.
\end{equation}
The multi-head outputs are concatenated and linearly projected to produce the final representation:
\begin{equation}
\operatorname{GMHSA}(\bm{Q}, \bm{K}, \bm{V}) = \operatorname{Concat}(\mathrm{head}_1, \ldots, \mathrm{head}_h) W^O,
\end{equation}
where $W^O\in\mathbb{R}^{C \times C}$ is the output projection matrix. $\operatorname{GMHA}(\cdot,\cdot,\cdot)$ denotes the global multi-head self-attention function. In our implementation $h = 12$, $C = 768$, and $d_h = C / h = 64$.

The key advantage of this design lies in the explicit injection of prompt tokens into the attention keys and values.  By doing so, the fine-grained local textures and boundary cues extracted by the CNN branch actively participate in global context modeling of the vision transformer while preserving the semantic integrity of the queries inherited from the pre-trained backbone.

\subsection{Boundary-Aware Adapter}
\label{subsec:Boundary-aware Adapter}
\begin{figure}[htbp]
\centering
\includegraphics[width=0.46\linewidth]{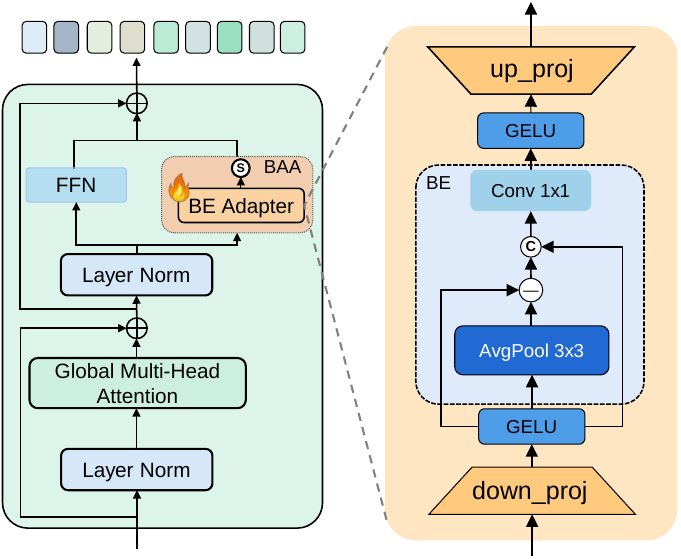}
\caption{Architecture of the proposed boundary-aware adapter (BAA). The BAA comprises a boundary enhancement (BE) module and a branch scaling factor $s$. The BE module acts as an explicit high-pass filter to compute and amplify the high-frequency components of the feature maps, enhancing the representation of fine-grained structures and boundaries. Concurrently, the scaling factor $s$ serves as a crucial regularizer, balancing the contribution of the high-frequency adapter branch against the primary frozen transformer backbone. In the diagram, \textcircled{c}, $\circleddash$, and \textcircled{s} denote feature concatenation, element-wise subtraction, and the scaling operation, respectively.}
\label{fig:adapter}  
\end{figure}

As shown in Fig.~\ref{fig:adapter}, the boundary-aware adapter (BAA) is deployed as a parallel branch alongside the original FFN, consisting of two essential modules: boundary enhancement (BE) and a branch scaling factor $s$.

\textbf{Boundary Enhancement}. Inspired by the Deep Edge Filter~\cite{lee2026deep}, we design the boundary enhancement (BE) module as an explicit high-pass filter constructed by subtracting low-pass filtered outputs from input features.  We implement the low-pass filtering via a 3$\times$3 average pooling operation. As depicted in Fig.~\ref{fig:adapter} (left), given the input features $\bm{X}_n \in \mathbb{R}^{N \times C}$ from layer normalization, we first project them into a low-dimensional bottleneck space via a linear down projection layer followed by a GELU activation:
\begin{equation}
\bm{F}_b = \mathrm{GELU}(\bm{X}_n \bm{W}_d), 
\end{equation}
where $\bm{F}_b \in \mathbb{R}^{N\times C} $ indicates the  bottleneck features, $\bm{W}_d \in \mathbb{R}^{C\times {d_r}}$ denotes the down projection matrix, and $d_r \ll C$ is the bottleneck dimension. Then, we employ a 3$\times$3 average pooling operation as the low-pass filter on $\bm{F}_b$ to yield the low-pass filtered results $\bm{F}_l$, and the generalizable, task-specific high-frequency representations $\bm{F}_h$ are isolated by subtracting $\bm{F}_l$ from $\bm{F}_b$:
\begin{equation}
\bm{F}_l = \mathrm{AvgPool}_{3\times3}(\bm{F}_b), \quad \bm{F}_h = \bm{F}_b - \bm{F}_l.
\end{equation}
Subsequently, the bottleneck feature $\bm{F}_b$ and the extracted boundary component $\bm{F}_h$ are concatenated along the channel dimension and fused through a 1$\times$1 convolution:
\begin{equation}
\bm{F}_f = \mathrm{Conv}_{1\times1}\left([\bm{F}_b; \bm{F}_h]\right).
\end{equation}
Finally, $\bm{F}_f$ is activated by GELU and projected back to the original dimension via an up-projection matrix $\bm{W}_u \in \mathbb{R}^{d_r \times C}$, producing the BE Adapter (denoted as BEA) output:
\begin{equation}
\mathrm{BEA}(\bm{X}_n) = \mathrm{GELU}(\bm{F}_f) \bm{W}_u.
\end{equation}

\textbf{Branch Scaling Factor $s$}. Following prior work on adapter-based fine-tuning~\cite{Chen2022AdaptFormer, he2021towards, jie2024convolutional} and branch scaling~\cite{yu2024branch}, we introduce a scaling factor $s$ to balance the contribution of the adapter branch relative to the frozen transformer pathway. The BE Adapter output is integrated into the FFN via element-wise summation:
\begin{equation}
\bm{X}^{\prime} = \mathrm{FFN}(\bm{X}_n) + s \cdot \mathrm{BEA}(\bm{X}_n),
\end{equation}
where $\mathrm{FFN}(\cdot)$ denotes the original feed-forward network and $\bm{X}^{\prime} $ serves as the input to subsequent transformer blocks.  As a hyperparameter, $s$ is helpful for the trade-off between the primary semantic pathway and the high-frequency adapter branch, and thus preventing optimization instability during early training stages. Our ablation studies demonstrate that setting $s=0.3$ yields optimal segmentation performance.

\subsection{Loss Function}
\label{subsec:loss_function}
We adopt a hybrid loss function that integrates both Dice loss ($\mathcal{L}_{dice}$) and binary cross-entropy (BCE) loss ($\mathcal{L}_{bce}$) to effectively address the issue of class imbalance in medical image segmentation. Let $N$ denote the number of pixels (\textit{i.e.}, $HW$), $g(i) \in (0,1)$ indicate the ground truth label, and $p(i) \in (0,1)$ represent the predicted probability for the foreground class.

The Dice loss evaluates the overlap between the predicted results and the ground truth labels, which is defined as:
\begin{equation}
\mathcal{L}_{dice} = 1 - \frac{2 \sum_{i=1}^{N} p(i) g(i)}{\sum_{i=1}^{N} p(i)^2 + \sum_{i=1}^{N} g(i)^2}.
\end{equation}
The BCE loss measures the pixel-wise classification error, and is defined as:
\begin{equation}
\mathcal{L}_{bce} = -\frac{0.85}{N} \sum_{i=1}^{N} \left( g(i) \cdot \log(p(i)) + (1-g(i)) \cdot \log(1-p(i)) \right).
\end{equation}

The overall hybrid loss function used for training is formulated as:
\begin{equation}
\mathcal{L} = \lambda \mathcal{L}_{dice} + (1-\lambda) \mathcal{L}_{bce},
\end{equation}
where $\lambda$ is a weighted factor that balances the impacts of Dice loss and BCE loss. In all our experiments, $\lambda$ is empirically set to $0.8$.

\section{Experimental Settings}
\label{sec:experimentalsets}
\subsection{Datasets}  
To comprehensively evaluate the performance and generalization capability of CVP-BAA, we conduct experiments on five publicly available datasets: BUSI~\cite{2020busi}, UDIAT~\cite{yap2020udiat}, TN3K~\cite{GONG2023106389TN3K}, CVC-ClinicDB~\cite{BERNAL2015CVC}, and ISIC-2018~\cite{Tschandl2018HAM10000}.
These datasets span three representative medical imaging modalities: ultrasound, endoscopy, and dermoscopy. Ultrasound imaging is particularly challenging due to inherent artifacts such as low contrast, speckle noise, and ambiguous lesion boundaries, making it a rigorous benchmark for assessing boundary-aware segmentation. Thus, we incorporate two breast ultrasound datasets (BUSI and UDIAT) alongside one thyroid ultrasound dataset (TN3K). Furthermore, to thoroughly assess cross-domain generalization, we include the CVC-ClinicDB dataset for endoscopy and the ISIC-2018 dataset for dermoscopy. This deliberately curated, multi-modality benchmark is designed to rigorously validate the robustness and broad clinical applicability of the proposed adaptation method.

\subsubsection{BUSI}
The BUSI dataset contains 780 breast ultrasound images with pixel-wise annotations, acquired using LOGIQ E9 and LOGIQ E9 Agile scanners. All the images have an average resolution of 500$\times$500 pixels in PNG format. To ensure fair comparison, we adopt the same data split as SAMUS~\cite{samus2024}, dividing the dataset into training, validation, and testing sets with a 7:1:2 ratio.

\subsubsection{UDIAT}
The UDIAT dataset comprises 163 pixel-wise annotated ultrasound images from female patients, each containing one or more lesions. The images are acquired using a Siemens ACUSON Sequoia C512 system with an 8.5 MHz 17L5 HD linear array transducer and have a mean resolution of 760$\times$570 pixels. For our experiments, the dataset is partitioned into 115 training, 16 validation, and 32 testing images.

\subsubsection{TN3K}
The TN3K dataset comprises 3,493 thyroid ultrasound images with pixel-wise nodule annotations, collected across three different ultrasound imaging systems: RESONA 70B, ARIETTA 850, and GE Logiq E9. Following the data partitioning protocol established by SAMUS~\cite{samus2024} and TRFE+~\cite{GONG2023106389TN3K}, we divide the dataset into 2,303 training, 576 validation, and 614 testing samples.

\subsubsection{CVC-ClinicDB}
The CVC-ClinicDB dataset consists of 612 colonoscopy images, originally utilized as the benchmark for the MICCAI 2015 Sub-Challenge on Automatic Polyp Detection. In our experiments, the dataset is randomly partitioned into 490 images for training, 61 for validation, and 61 for testing.

\subsubsection{ISIC-2018}
The ISIC-2018 dataset is widely recognized for benchmarking skin lesion segmentation methods. It comprises 2,594 clinical dermoscopy images with varying resolutions, which are randomly partitioned into 1,868 training images, 467 validation images, and 259 testing images, respectively.

\subsection{Evaluation Metrics}
To ensure a fair comparison with state-of-the-art methods and to comprehensively evaluate the performance of CVP-BAA, we employ six standard segmentation metrics: Dice coefficient (Dice), Hausdorff distance (HD), mean Intersection over Union (mIoU), Accuracy (Acc), Precision (Prec), and Recall (Rec). For the ultrasound datasets: BUSI, UDIAT, and TN3K, we report Dice and HD; for ISIC-2018 and CVC-ClinicDB, we report Dice, mIoU, Acc, Prec, and Rec. Additionally, we assess computational efficiency through the number of parameters (\#Params), floating-point operations (FLOPs), and inference speed (FPS). All results are reported as mean$\pm$standard deviation over three independent runs with different random seeds to verify the robustness and generalizability of the proposed CVP-BAA.

\subsection{Implementation Details} 
We adopt the pre-trained SAM with a ViT-B backbone, originally trained on the SA-1B dataset, as our foundational architecture. 
All three components in this version of SAM, namely, the image encoder, prompt encoder, and mask decoder, remain frozen during training.
Only the parameters of the parallel CNN branch, along with the CVP and BAA components, are optimized. All experiments are conducted on an NVIDIA GeForce RTX 3090 GPU with 24 GB of memory. The models are trained for 200 epochs using the Adam optimizer with an initial learning rate of $1e-4$ and a batch size of 8. The training objective is supervised by a hybrid loss function combining Dice loss and binary cross-entropy (BCE) loss. In addition, following the protocol of SAMUS~\cite{samus2024}, we employ a single-point prompt for interaction, and all input images are uniformly resized to 256$\times$256 pixels.

\section{Experimental Results}
\subsection{Comparisons with other methods}
\label{quantitativeanalysis}
We evaluate CVP-BAA against a comprehensive suite of state-of-the-art baselines. This includes representative task-specific architectures spanning CNN-based models like UNet~\cite{unet2015} and AAU-Net~\cite{aaunet2022}, pure transformer frameworks such as Swin-Unet~\cite{swinunet2022} and MISSFormer~\cite{missformer2021}, and hybrid CNN-Transformer models including TransFuse~\cite{transfuse2021}, H2Former~\cite{h2former2023}, and BRAU-Net++~\cite{lan2024brau}. Additionally, we benchmark our method against leading SAM adaptation strategies, namely vanilla SAM~\cite{sam2023}, MedSAM~\cite{Medsam2024}, SAMed~\cite{SAMed2023}, and Med-SA~\cite{Wu2025Med-SA}.

\subsubsection{Comparison on BUSI, UDIAT, and TN3K}
\textbf{Quantitative Results}. 
Table~\ref{tab:compression_busi_tn3k_udiat} summarizes the quantitative comparisons between CVP-BAA, representative task-specific segmentation models, and recent SAM adaptations on three ultrasound datasets: BUSI, UDIAT, and TN3K, with results reported as mean$\pm$standard deviation.
Overall, CVP-BAA consistently achieves the highest Dice scores across all three benchmarks, reaching 86.71\%, 89.72\%, and 85.23\% on BUSI, UDIAT, and TN3K, respectively. Furthermore, in terms of boundary precision, our method yields the lowest HD values, recording 24.52 mm, 13.17 mm, and 27.77 mm for the respective datasets, demonstrating superior boundary delineation. These results confirm that CVP-BAA effectively adapts SAM for challenging medical image segmentation tasks, delivering both robust region coverage and precise boundary accuracy.

As detailed in Table~\ref{tab:compression_busi_tn3k_udiat}, CVP-BAA establishes a new state-of-the-art on the BUSI dataset, consistently outperforming all baseline categories, including CNN-based, Transformer-based, hybrid CNN-Transformer, and existing SAM adaptation methods, across both Dice and HD metrics. While task-specific hybrid architectures such as H2Former and BRAU-Net++ achieve competitive results by combining the local feature extraction of CNNs with the global context modeling of Transformers, they remain suboptimal compared to advanced parameter-efficient fine-tuning (PEFT) strategies. We attribute the superior performance of CVP-BAA to its dual-mechanism design: it effectively harnesses the rich, generalized representations of the pre-trained foundation model while explicitly injecting local, image-specific inductive biases via our convolutional visual prompting (CVP) strategy and simultaneously improves boundary delineation through the boundary-aware adapter (BAA). 
Furthermore, it is worth noting that while vanilla SAM~\cite{sam2023} demonstrates remarkable zero-shot generalization on natural images from the SA-1B dataset, it suffers from severe performance degradation when directly applied to medical images. This stark contrast underscores the substantial domain shift between the two domains, thereby validating the necessity of our proposed adaptation framework.

\begin{table}[htbp]
\centering
\footnotesize
\caption{Quantitative comparison (mean\mypm{std}) of the proposed CVP-BAA against state-of-the-art task-specific segmentation models and SAM adaptation methods on the BUSI, UDIAT, and TN3K datasets. Performance is evaluated using the Dice score (\%) and Hausdorff distance (HD). The symbol $\uparrow$ indicates the larger the better. The symbol $\downarrow$ denotes the smaller the better. The best results are marked in \textbf{bold}, and the second-best results are \underline{underlined}. Notably, the results for the vanilla SAM highlighted in gray are obtained via direct zero-shot inference, without any domain-specific fine-tuning or adaptation. As observed, CVP-BAA consistently outperforms all competing methods across all evaluation metrics.}
\label{tab:compression_busi_tn3k_udiat}
\begin{tabular}{@{}l|cc|cc|cc}
\hline
\multirow{2}{*}{Methods} & \multicolumn{2}{c|}{BUSI} & \multicolumn{2}{c|}{UDIAT} & \multicolumn{2}{c}{TN3K} \\
\cline{2-7}
& Dice (\%) $\uparrow$ & HD (mm) $\downarrow$ & Dice (\%) $\uparrow$ & HD (mm) $\downarrow$  & Dice (\%) $\uparrow$ & HD (mm) $\downarrow$\\
\hline
U-Net~\cite{unet2015}   & 77.34\mypm{1.14}   & 34.31\mypm{1.38}   & 83.88\mypm{0.37}   & 19.75\mypm{1.48}  & 79.01\mypm{0.01}   & 33.76\mypm{0.35} \\
AAU-Net~\cite{aaunet2022}        & 80.62\mypm{0.18}   & 29.32\mypm{0.56}     & 82.78\mypm{0.53}   & 16.28\mypm{1.17} & 82.12\mypm{0.15}   & 30.87\mypm{0.69}\\
Swin-Unet~\cite{swinunet2022}     & 67.70\mypm{0.55}   & 46.85\mypm{0.37}    & 64.62\mypm{0.81}   & 27.76\mypm{0.38}  & 70.51\mypm{0.54}   & 43.6\mypm{0.51}\\
MISSFormer~\cite{missformer2021} & 78.83\mypm{0.46}   & 32.29\mypm{0.95}    & 81.47\mypm{1.01}   & 16.41\mypm{1.28}  & 79.94\mypm{0.54}   & 32.48\mypm{0.51}\\
TransFuse~\cite{transfuse2021}   & 74.23\mypm{1.09}   & 34.03\mypm{1.48}     & 71.35\mypm{1.79}   & 26.23\mypm{0.75} & 78.58\mypm{0.40}   & 32.47\mypm{0.42}\\
H2Former~\cite{h2former2023}     & 81.61\mypm{0.43}   & 28.06\mypm{0.83}     & 84.92\mypm{0.67}   & 19.84\mypm{1.56} & 81.90\mypm{0.49}   & 30.56\mypm{0.03}\\
BRAU-Net++~\cite{lan2024brau}     & 80.76\mypm{1.12}   & 30.95\mypm{1.83}    & 83.04\mypm{0.96}   & 18.22\mypm{0.56}  & 80.70\mypm{0.17}   & 32.25\mypm{0.36}\\
\hline
SAM~\cite{sam2023} & \cellcolor{lightgray}{54.01\mypm{0.00}} & \cellcolor{lightgray}{82.39\mypm{0.00}}   & \cellcolor{lightgray}{42.95\mypm{0.00}}   & \cellcolor{lightgray}{121.09\mypm{0.00}} & \cellcolor{lightgray}{29.59\mypm{0.00}} & \cellcolor{lightgray}{134.87\mypm{0.00}}\\
MedSAM~\cite{Medsam2024}   & 81.85\mypm{0.23}  & 30.67\mypm{0.42}    & 84.80\mypm{0.63}   & 19.13\mypm{1.43} & 75.87\mypm{0.21}  & 35.42\mypm{0.32}\\
SAMed~\cite{SAMed2023}  & 83.47\mypm{0.34} & 29.98\mypm{0.64}   & 86.25\mypm{0.39}   & 28.81\mypm{0.64} & 82.39\mypm{0.34} & 35.16\mypm{0.28}\\
Med-SA~\cite{Wu2025Med-SA}  & 83.72\mypm{0.41} & 27.83\mypm{0.49}  & 87.56\mypm{0.23}   & 18.15\mypm{1.32}  & 83.17\mypm{0.29} & 29.40\mypm{0.25}\\
SAMUS~\cite{samus2024} & \underline{85.07\mypm{0.46}} & \underline{25.85\mypm{0.71}}  & \underline{89.43\mypm{0.25}}   & \underline{14.09\mypm{0.36}} & \underline{84.46\mypm{0.21}} & \underline{28.41\mypm{0.18}}\\
\hline
\textbf{CVP-BAA} & \textbf{86.71\mypm{0.33}} & \textbf{24.52\mypm{0.44}}  & \textbf{89.72\mypm{0.23}}   & \textbf{13.17\mypm{0.58}} & \textbf{85.23\mypm{0.30}} & \textbf{27.77\mypm{0.42}}\\
\hline
\end{tabular}%
\end{table}

\begin{figure}[htbp]
    \imagestyle{1pt}{0.1}
     \newcommand{\imgw}{0.121\linewidth}
   
    % ===================== (a) task-specific models =====================
    \centering
    \begin{tabular}{ccccccccc}
       \footnotesize
        % row 1 - BUSI
        \rotatebox[origin=c]{90}{BUSI}&
        \includegraphics[width=\imgw, valign=c]{image-malignant0171.png} &
        \includegraphics[width=\imgw, valign=c]{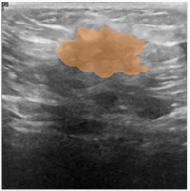}  &
        \includegraphics[width=\imgw, valign=c]{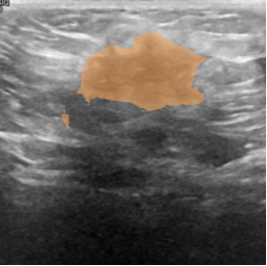} &
        \includegraphics[width=\imgw, valign=c]{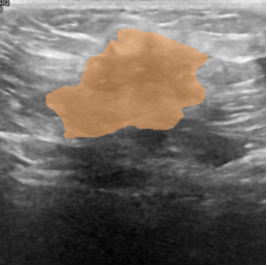} &
        \includegraphics[width=\imgw, valign=c]{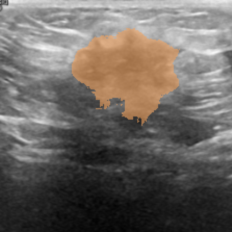} &
        \includegraphics[width=\imgw, valign=c]{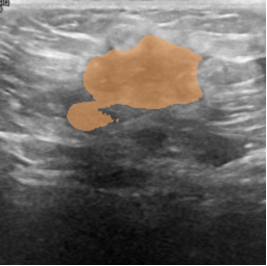} &
        \includegraphics[width=\imgw, valign=c]{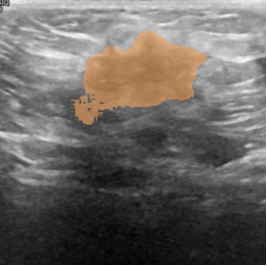} &
        \includegraphics[width=\imgw, valign=c]{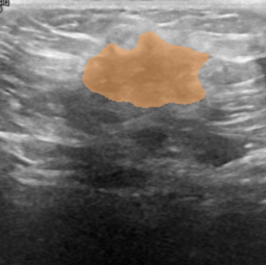} \\[7mm]
         % row 2 - UDIAT
         \rotatebox[origin=c]{90}{UDIAT}&
        \includegraphics[width=\imgw, valign=c]{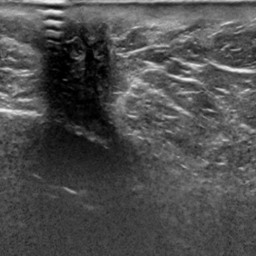} &
        \includegraphics[width=\imgw, valign=c]{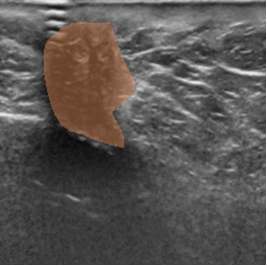}   &
        \includegraphics[width=\imgw, valign=c]{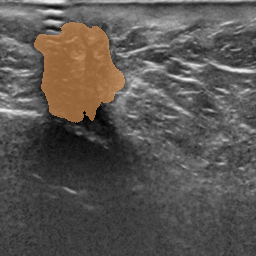} &
        \includegraphics[width=\imgw, valign=c]{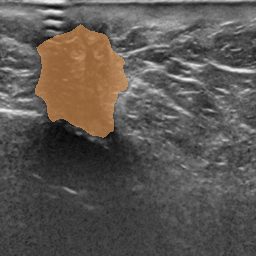} &
        \includegraphics[width=\imgw, valign=c]{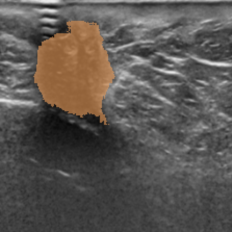} &
        \includegraphics[width=\imgw, valign=c]{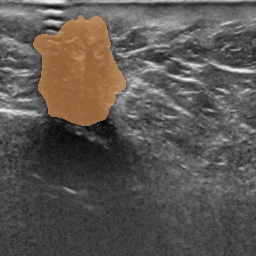} &
        \includegraphics[width=\imgw, valign=c]{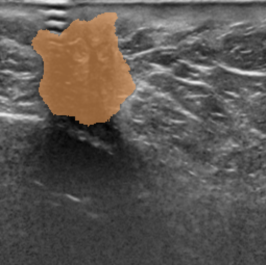} &
        \includegraphics[width=\imgw, valign=c]{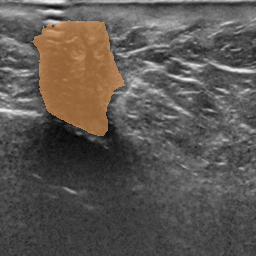}  \\[7mm]
        % row 3 - TN3K
        \rotatebox[origin=c]{90}{TN3K}&
        \includegraphics[width=\imgw, valign=c]{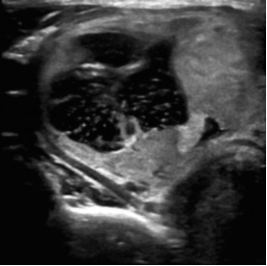} &
        \includegraphics[width=\imgw, valign=c]{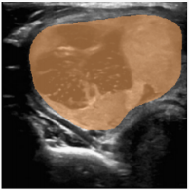}    &
        \includegraphics[width=\imgw, valign=c]{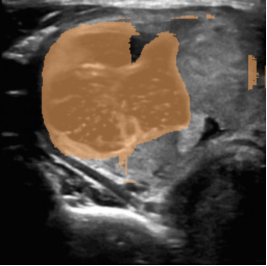} &
        \includegraphics[width=\imgw, valign=c]{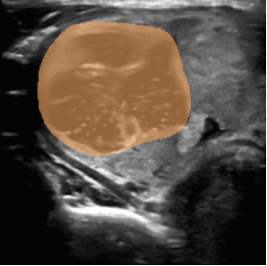} &
        \includegraphics[width=\imgw, valign=c]{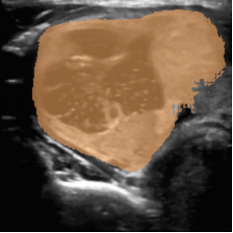} &
        \includegraphics[width=\imgw, valign=c]{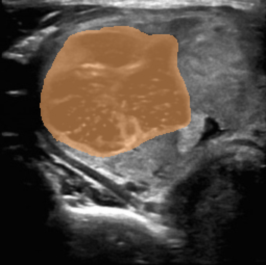} &
        \includegraphics[width=\imgw, valign=c]{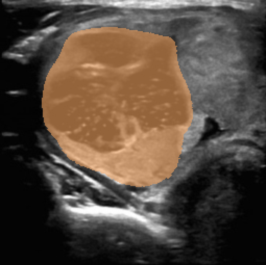}&
        \includegraphics[width=\imgw, valign=c]{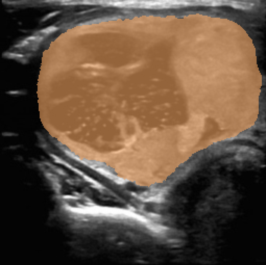} \\[7mm]

        &
        {\scriptsize Image} &
        {\scriptsize GT} &
        {\scriptsize UNet} &
        {\scriptsize AAU-Net} &
        {\scriptsize MISSFormer} &
        {\scriptsize H2Former} &
        {\scriptsize BRAU-Net++} &
        {\scriptsize CVP-BAA}  \\

        \multicolumn{9}{c}{\textbf{(a)}}\\[1mm]
        \rotatebox[origin=c]{90}{BUSI}&
        \includegraphics[width=\imgw, valign=c]{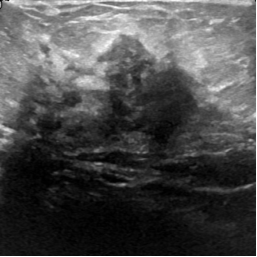} &
        \includegraphics[width=\imgw, valign=c]{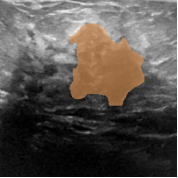}  &
        \includegraphics[width=\imgw, valign=c]{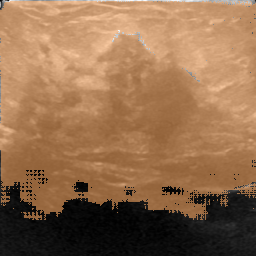} &
        \includegraphics[width=\imgw, valign=c]{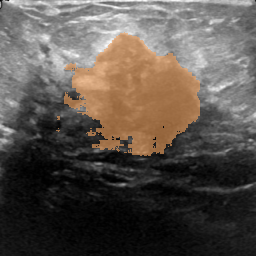} &
        \includegraphics[width=\imgw, valign=c]{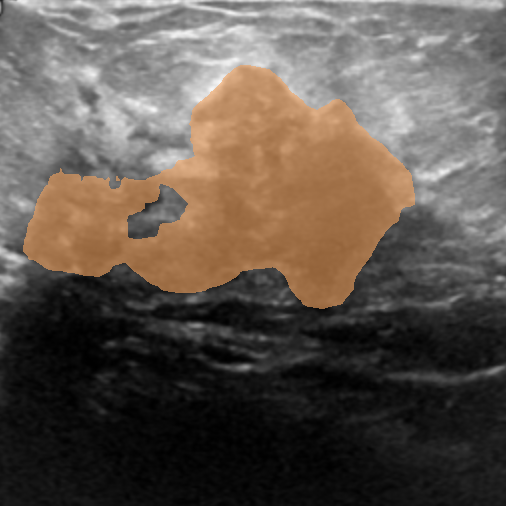} &
        \includegraphics[width=\imgw, valign=c]{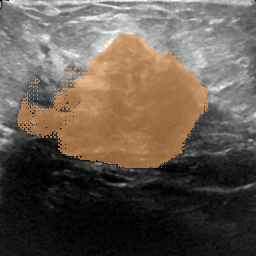} &
        \includegraphics[width=\imgw, valign=c]{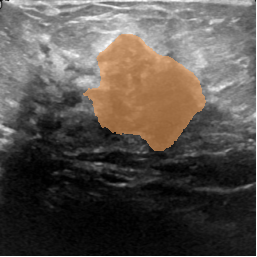} &
        \includegraphics[width=\imgw, valign=c]{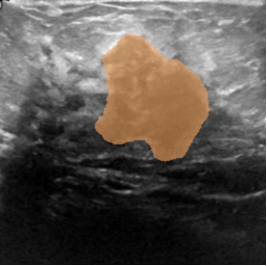} \\[7mm]
        
        \rotatebox[origin=c]{90}{UDIAT}&
        \includegraphics[width=\imgw, valign=c]{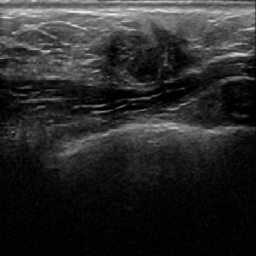} &
        \includegraphics[width=\imgw, valign=c]{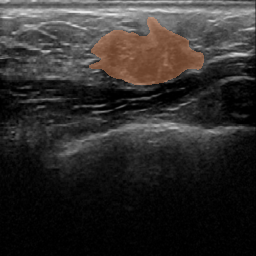} &
        \includegraphics[width=\imgw, valign=c]{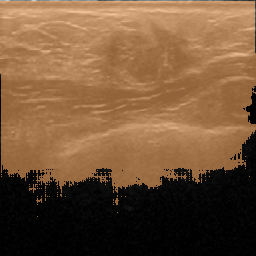} &
        \includegraphics[width=\imgw, valign=c]{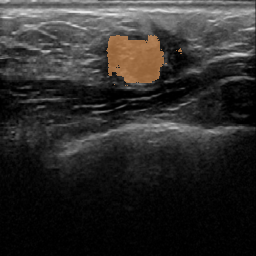} &
        \includegraphics[width=\imgw, valign=c]{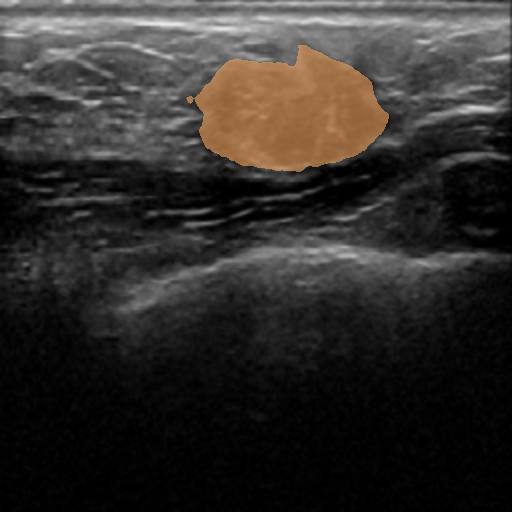} &
        \includegraphics[width=\imgw, valign=c]{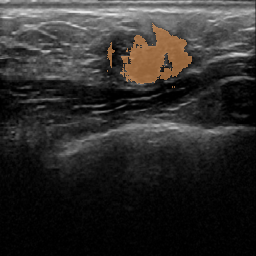} &
        \includegraphics[width=\imgw, valign=c]{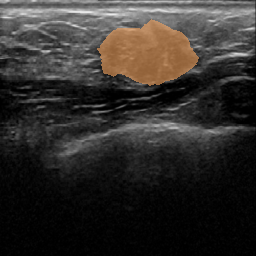} &
        \includegraphics[width=\imgw, valign=c]{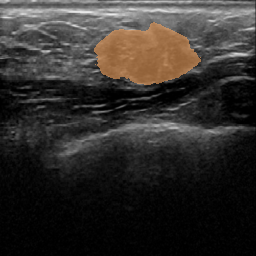} \\[7mm]

       \rotatebox[origin=c]{90}{TN3K}&
        \includegraphics[width=\imgw, valign=c]{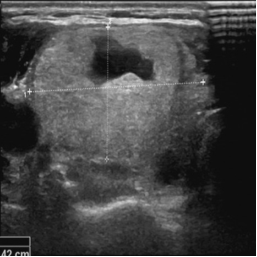} &
        \includegraphics[width=\imgw, valign=c]{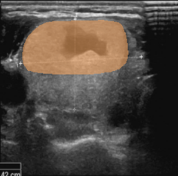} &
        \includegraphics[width=\imgw, valign=c]{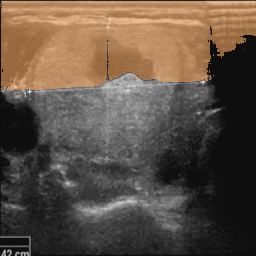} &
        \includegraphics[width=\imgw, valign=c]{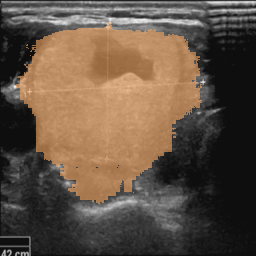} &
        \includegraphics[width=\imgw, valign=c]{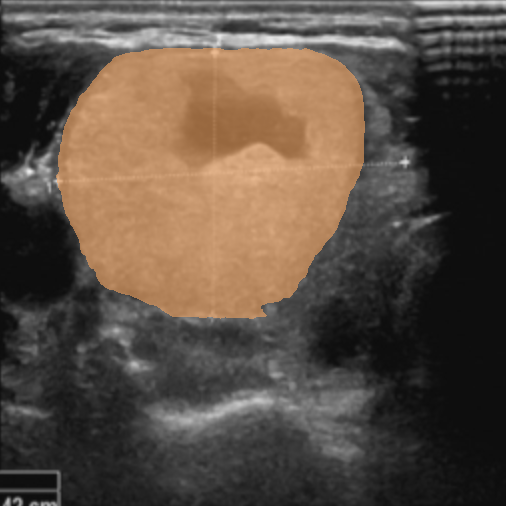} &
        \includegraphics[width=\imgw, valign=c]{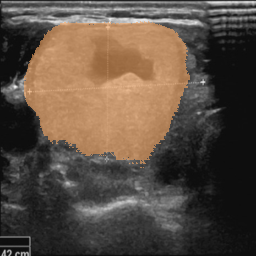} &
        \includegraphics[width=\imgw, valign=c]{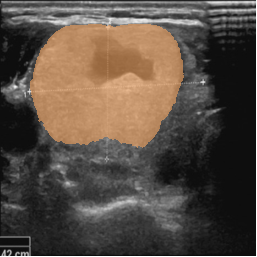} &
        \includegraphics[width=\imgw, valign=c]{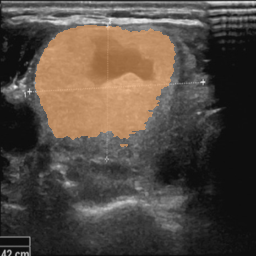} \\[7mm]
        
        &
        {\scriptsize Image} &
        {\scriptsize GT  } &
        {\scriptsize SAM }&
        {\scriptsize MedSAM} &
        {\scriptsize SAMed}&
        {\scriptsize Med-SA} &
        {\scriptsize SAMUS} &
        {\scriptsize CVP-BAA}\\
        \multicolumn{9}{c}{\textbf{(b)}}\\  
    \end{tabular}
    \caption{Qualitative comparison of segmentation results on the BUSI, UDIAT, and TN3K datasets. The proposed CVP-BAA is compared against: (a) task-specific segmentation models, including UNet~\cite{unet2015}, AAU-Net~\cite{aaunet2022}, MISSFormer~\cite{missformer2021}, H2Former~\cite{h2former2023}, and BRAU-Net++~\cite{lan2024brau}; and (b) SAM and its adaptation methods, namely SAM~\cite{sam2023}, MedSAM~\cite{Medsam2024}, SAMed~\cite{SAMed2023}, Med-SA~\cite{Wu2025Med-SA}, and SAMUS~\cite{samus2024}. As illustrated, CVP-BAA produces lesion masks and boundary contours that most closely match the ground truth.
    }
    \label{fig:multi_dataset}
\end{figure}

\textbf{Qualitative Results}. Fig.~\ref{fig:multi_dataset} presents the qualitative segmentation results across the BUSI, UDIAT, and TN3K datasets. As observed, while recent SAM-adapted methods mitigate some baseline limitations, both task-specific models and existing SAM adaptations still struggle to accurately delineate boundaries in complex ultrasound images. These methods frequently exhibit blurred edges and over- or under-segmentation in challenging cases, leading to extensive false positives and negatives. In contrast, CVP-BAA consistently generates highly accurate masks with clearer boundaries and fewer segmentation errors. This superior performance could stem from the robust foundational representations of SAM, effectively leveraged and refined by our CVP and BAA modules, which explicitly inject local structural priors and enhance high-frequency boundary details.

\subsubsection{Comparison on CVC-ClinicDB}
\textbf{Quantitative Results}. 
The quantitative evaluation on the CVC-ClinicDB dataset is reported in Table~\ref{tab:cvc}. Our proposed CVP-BAA achieves state-of-the-art performance across all evaluation metrics, yielding a Dice score of 94.11\%, IoU of 89.36\%, Accuracy of 99.16\%, Precision of 94.57\%, and Recall of 94.21\%. Notably, compared to the strongest SAM-based competitor, SAMUS~\cite{samus2024}, CVP-BAA delivers consistent improvements with margins of +0.83\% in Dice, +1.27\% in mIoU, +0.15\% in Accuracy, +1.46\% in Precision, and +0.02\% in Recall. Among task-specific approaches, BRAU-Net++~\cite{lan2024brau} achieves the highest baseline performance, with 92.18\% Dice, 88.21\% IoU, and 93.19\% Recall. However, CVP-BAA still substantially outperforms it.

\begin{table}[htbp]
\centering
\footnotesize
\caption{Quantitative comparison (mean\mypm{std}) of the proposed CVP-BAA against state-of-the-art task-specific segmentation models and SAM adaptation methods on the CVC-ClinicDB dataset. The symbol $\uparrow$ indicates the larger the better. The best results are marked in \textbf{bold}, and the second-best results are \underline{underlined}. The results for the vanilla SAM highlighted in gray are obtained via direct zero-shot inference, without any domain-specific fine-tuning or adaptation.}
\label{tab:cvc}
\begin{tabular}{@{}l|ccccc}
\hline
Methods\rule{0pt}{2.6ex} & Dice (\%) $\uparrow$ & mIoU (\%) $\uparrow$ & Acc (\%) $\uparrow$ & Prec (\%) $\uparrow$ & Rec (\%) $\uparrow$ \\
\hline
U-Net~\cite{unet2015}              & 88.27\mypm{1.02} & 82.37\mypm{1.47} & 98.44\mypm{0.02} & 89.28\mypm{1.46} & 89.62\mypm{0.63} \\
AAU-Net~\cite{aaunet2022}          & 92.08\mypm{0.70} & 87.24\mypm{0.81} & 98.96\mypm{0.08} & \underline{94.20\mypm{1.23}} & 92.07\mypm{0.80} \\
Swin-Unet~\cite{swinunet2022}       & 87.77\mypm{0.46} & 84.14\mypm{0.71} & 95.22\mypm{0.45} & 89.35\mypm{1.10} & 90.49\mypm{0.57} \\
MISSFormer~\cite{missformer2021}   & 91.24\mypm{0.36} & 85.39\mypm{0.71} & 95.22\mypm{0.45} & 89.35\mypm{1.11} & 90.49\mypm{0.57} \\
TransFuse~\cite{transfuse2021}     & 85.01\mypm{0.90} & 78.73\mypm{0.92} & 98.13\mypm{0.10} & 87.61\mypm{1.23} & 85.56\mypm{0.55} \\
H2Former~\cite{h2former2023}       & 91.21\mypm{0.31} & 85.36\mypm{0.33} & 98.76\mypm{0.11} & 92.58\mypm{1.43} & 92.03\mypm{1.18} \\
BRAU-Net++~\cite{lan2024brau}       & 92.18\mypm{0.25} & 88.21\mypm{0.22} & 98.94\mypm{0.10} & 93.74\mypm{0.12} & 93.19\mypm{0.67} \\
\hline
SAM~\cite{sam2023}                 & \cellcolor{lightgray}{59.73\mypm{0.00}} & \cellcolor{lightgray}{50.02\mypm{0.00}} & \cellcolor{lightgray}{84.19\mypm{0.00}} & \cellcolor{lightgray}{58.00\mypm{0.00}} & \cellcolor{lightgray}{87.01\mypm{0.00}} \\
MedSAM~\cite{Medsam2024}         & 87.62\mypm{1.83} & 80.74\mypm{1.25} & 97.99\mypm{0.25} & 89.02\mypm{1.69} & 88.51\mypm{1.84}  \\
SAMed~\cite{SAMed2023}             & \underline{93.38\mypm{0.59}} & \underline{88.60\mypm{0.79}} & \underline{99.06\mypm{0.12}} & 93.49\mypm{0.24} & 93.20\mypm{0.48} \\
Med-SA~\cite{Wu2025Med-SA}              & 93.21\mypm{0.34} & 88.56\mypm{0.45} & 98.98\mypm{0.03} & 93.01\mypm{0.56} & 93.97\mypm{0.32} \\
SAMUS~\cite{samus2024}             & 93.28\mypm{0.48} & 88.09\mypm{0.39} & 99.01\mypm{0.14} & 93.11\mypm{0.63} & \underline{94.19\mypm{0.38}} \\
\hline
CVP-BAA                             & \textbf{94.11\mypm{0.21}} & \textbf{89.36\mypm{0.20}} & \textbf{99.16\mypm{0.02}} & \textbf{94.57\mypm{0.35}} & \textbf{94.21\mypm{0.42}} \\
\hline
\end{tabular}%
\end{table}

Regarding foundation models, vanilla SAM exhibits severe performance degradation due to the inherent domain shift between natural and medical images. Nevertheless, advanced parameter-efficient fine-tuning (PEFT) methods like SAMed, SAMUS, and CVP-BAA can recover this performance drop and even surpass traditional task-specific models. This performance hierarchy highlights a critical insight: simply applying SAM is insufficient due to the domain shift, but strategically adapting it yields superior results. CVP-BAA excels by explicitly injecting local CNN-derived structural priors into the global attention mechanism and amplifying high-frequency boundary cues, thereby achieving highly accurate and robust delineation of complex polyp structures.

\begin{figure}[htbp]
    \centering
    \footnotesize
   \imagestyle{0.96pt}{0.85}
     \newcommand{\imgw}{0.12\linewidth}

    \begin{tabular}{cccccccc}
        % row 1 (213)
        \includegraphics[width=\imgw]{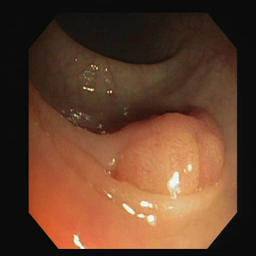} &
        \includegraphics[width=\imgw]{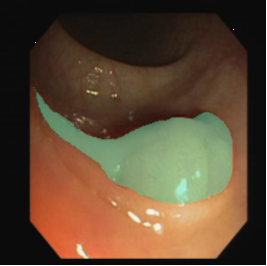} &
        \includegraphics[width=\imgw]{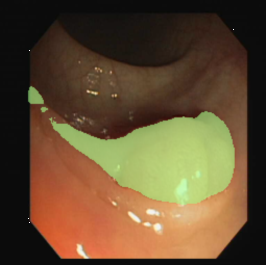} &
        \includegraphics[width=\imgw]{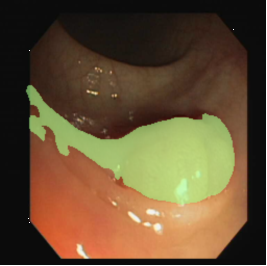} &
        \includegraphics[width=\imgw]{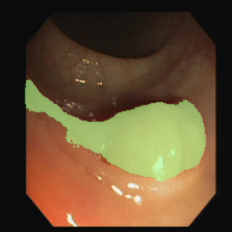} &
        \includegraphics[width=\imgw]{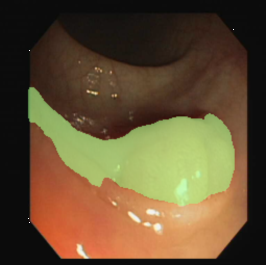} &
        \includegraphics[width=\imgw]{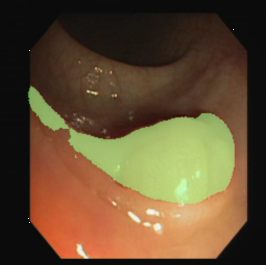} &
        \includegraphics[width=\imgw]{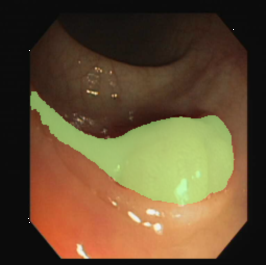}\\[-0.5mm]

        % row 2 (432)
        \includegraphics[width=\imgw]{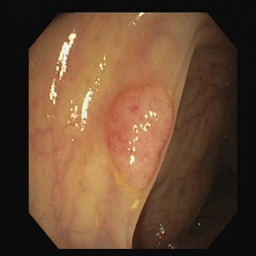} &
        \includegraphics[width=\imgw]{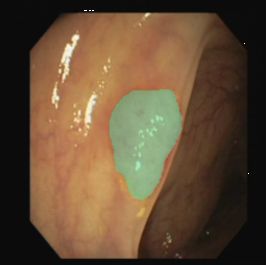}   &
        \includegraphics[width=\imgw]{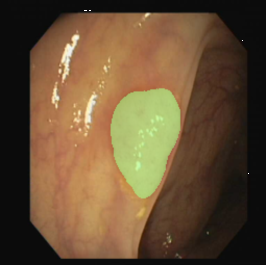} &
        \includegraphics[width=\imgw]{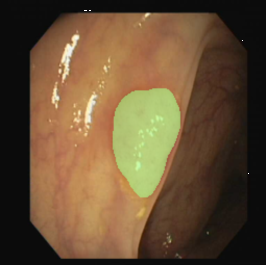} &
        \includegraphics[width=\imgw]{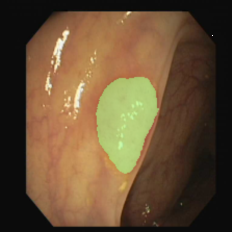} &
        \includegraphics[width=\imgw]{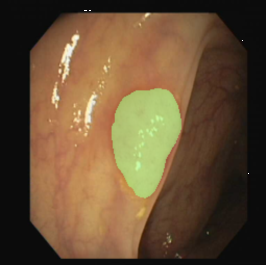} &
        \includegraphics[width=\imgw]{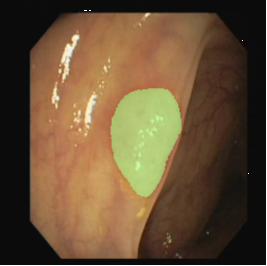} &
        \includegraphics[width=\imgw]{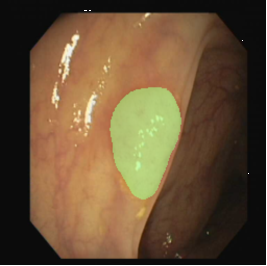} \\[-0.5mm]

        % row 3 (301)
        \includegraphics[width=\imgw]{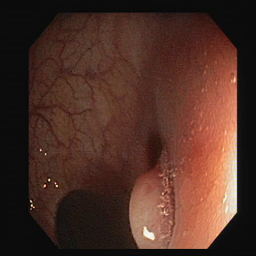} &
        \includegraphics[width=\imgw]{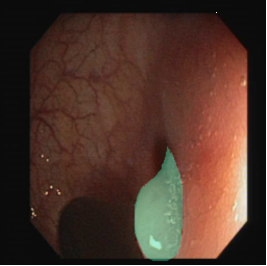} &
        \includegraphics[width=\imgw]{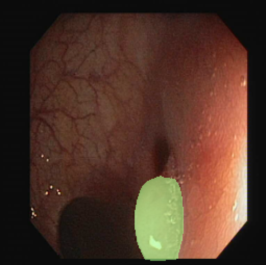} &
        \includegraphics[width=\imgw]{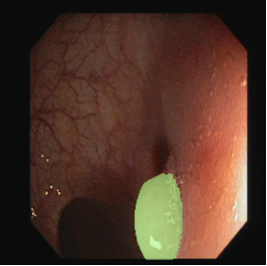} &
        \includegraphics[width=\imgw]{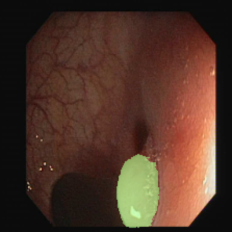} &
        \includegraphics[width=\imgw]{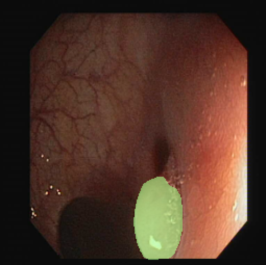} &
        \includegraphics[width=\imgw]{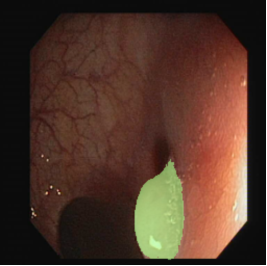} &
        \includegraphics[width=\imgw]{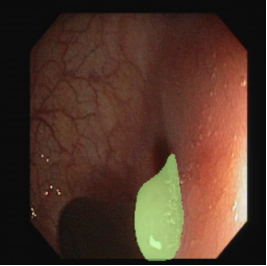}  \\
        
        Image &
        GT &
        UNet&
        AAU-Net &
        MISSFormer &
        H2Former &
        BRAU-Net++ &
        CVP-BAA  \\
       \multicolumn{8}{c}{\textbf{(a)}}\\[1mm]

       \includegraphics[width=\imgw]{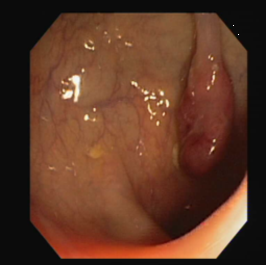} &
        \includegraphics[width=\imgw]{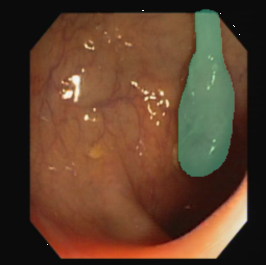} &
        \includegraphics[width=\imgw]{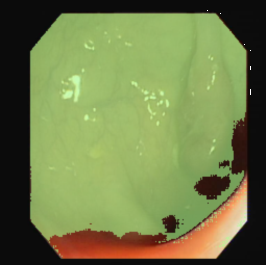} &
        \includegraphics[width=\imgw]{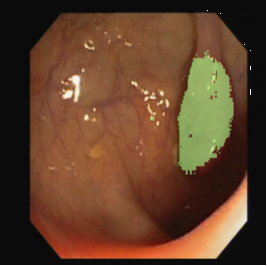} &
        \includegraphics[width=\imgw]{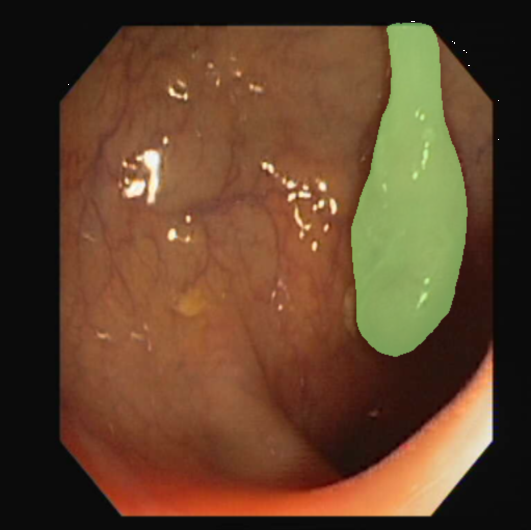} &
        \includegraphics[width=\imgw]{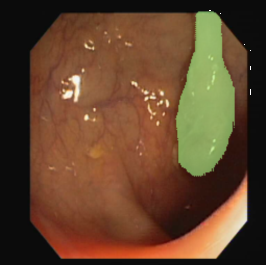} &
        \includegraphics[width=\imgw]{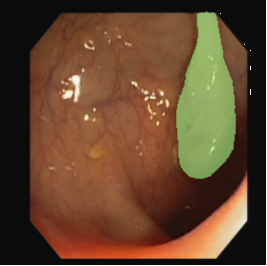} &
        \includegraphics[width=\imgw]{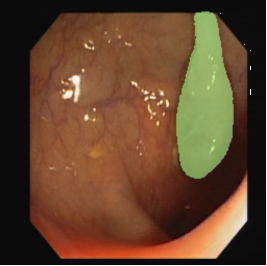} \\[-0.5mm]

        % row 2
        \includegraphics[width=\imgw]{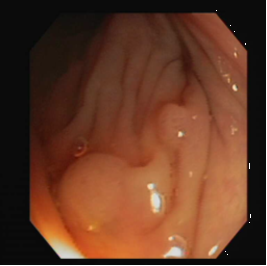} &
        \includegraphics[width=\imgw]{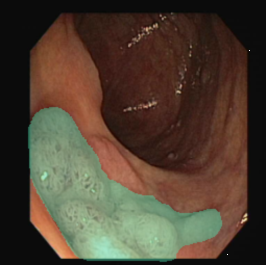}  &
        \includegraphics[width=\imgw]{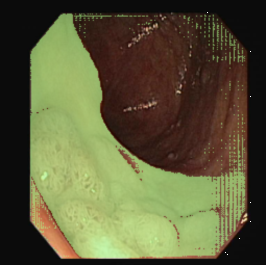} &
        \includegraphics[width=\imgw]{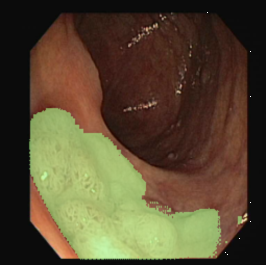} &
        \includegraphics[width=\imgw]{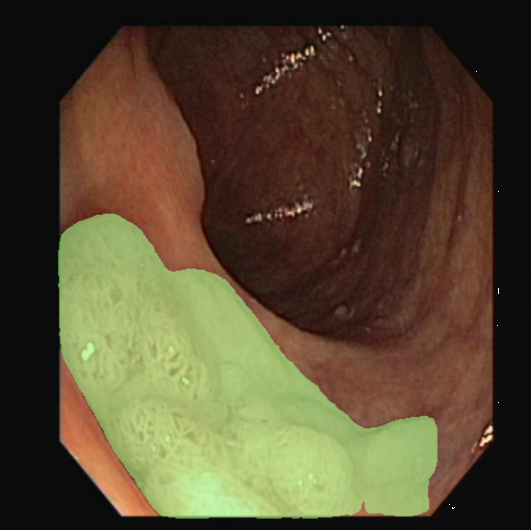} &
        \includegraphics[width=\imgw]{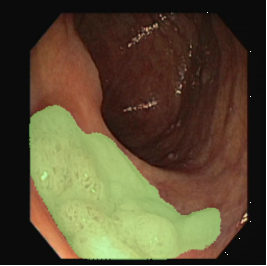} &
        \includegraphics[width=\imgw]{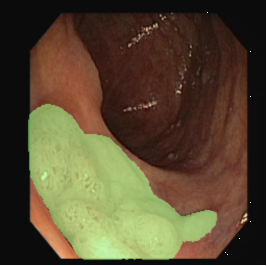} &
        \includegraphics[width=\imgw]{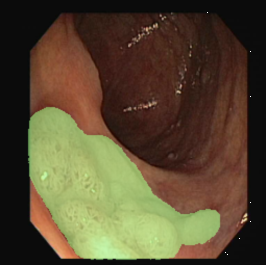}\\[-0.5mm]

        % row 3
        \includegraphics[width=\imgw]{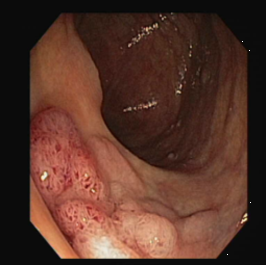} &
        \includegraphics[width=\imgw]{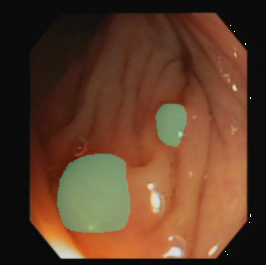} &
        \includegraphics[width=\imgw]{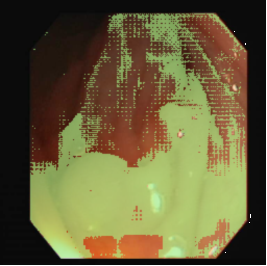} &
        \includegraphics[width=\imgw]{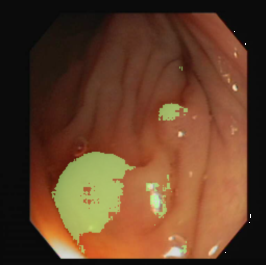} &
        \includegraphics[width=\imgw]{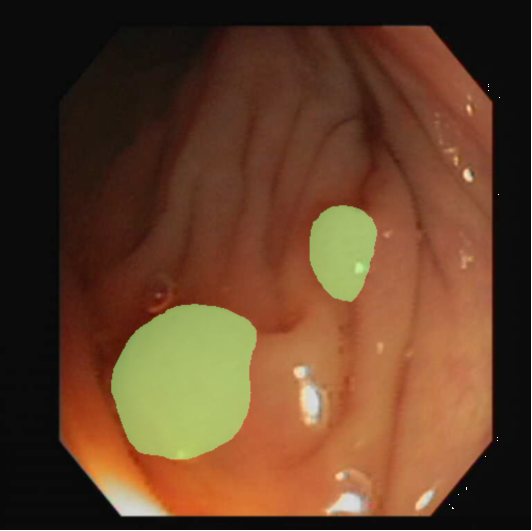} &
        \includegraphics[width=\imgw]{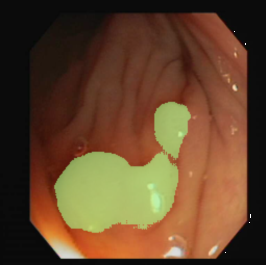} &
        \includegraphics[width=\imgw]{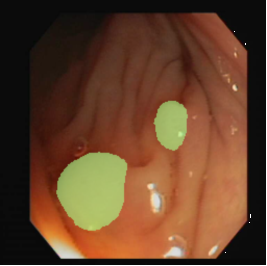} &
        \includegraphics[width=\imgw]{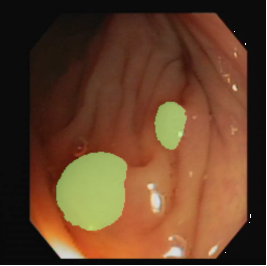} \\
       
        Image &
        GT &
        SAM &
        MedSAM &
        SAMed &
        Med-SA &
        SAMUS &
        CVP-BAA  \\
        \multicolumn{8}{c}{\textbf{(b)}}\\ 
    \end{tabular}
    \caption{Qualitative comparison of segmentation results on the CVC-ClinicDB dataset. The proposed CVP-BAA is compared against: (a) task-specific segmentation models, including UNet~\cite{unet2015}, AAU-Net~\cite{aaunet2022}, MISSFormer~\cite{missformer2021}, H2Former~\cite{h2former2023}, and BRAU-Net++~\cite{lan2024brau}; and (b) SAM and its adaptation methods, namely SAM~\cite{sam2023}, MedSAM~\cite{Medsam2024}, SAMed~\cite{SAMed2023}, Med-SA~\cite{Wu2025Med-SA}, and SAMUS~\cite{samus2024}. 
    Visually, CVP-BAA yields lesion and boundary delineations that more accurately approximate the ground truth.
    }
    \label{fig:cvc_task_foundation}
\end{figure}

\textbf{Qualitative Results}. 
The visual comparisons in Fig.~\ref{fig:cvc_task_foundation} show that the segmentation results generated by CVP-BAA closely match the ground truth. These qualitative findings underscore the robustness and high reliability of CVP-BAA in tackling complex medical image segmentation tasks, further supporting the quantitative results.

\begin{table}[htbp]
\centering
\footnotesize
\caption{Quantitative comparison (mean\mypm{std}) of the proposed CVP-BAA against state-of-the-art task-specific segmentation models and SAM adaptation methods on the ISIC-2018 dataset. The symbol $\uparrow$ indicates the larger the better. The best results are marked in \textbf{bold}, and the second-best results are \underline{underlined}. The results for the vanilla SAM highlighted in gray are obtained via direct zero-shot inference, without any domain-specific fine-tuning or adaptation.}
\label{tab:isic}
\begin{tabular}{@{}l|ccccc}
\hline
Methods & Dice (\%) $\uparrow$ & mIoU (\%) $\uparrow$ & Acc (\%) $\uparrow$ & Prec (\%) $\uparrow$ & Rec (\%) $\uparrow$ \\
\hline
U-Net~\cite{unet2015}            & 87.95\mypm{0.21} & 80.68\mypm{0.26} & 95.13\mypm{0.10} & 90.34\mypm{0.54} & 88.77\mypm{0.47} \\
AAU-Net~\cite{aaunet2022}        & 88.85\mypm{0.50} & 81.98\mypm{0.63} & 95.42\mypm{0.36} & 91.32\mypm{1.24} & 89.85\mypm{0.71} \\
Swin-Unet~\cite{swinunet2022}     & 87.35\mypm{0.07} & 80.57\mypm{1.22} & 95.12\mypm{0.33} & 89.84\mypm{1.36} & 90.62\mypm{0.63} \\
MISSFormer~\cite{missformer2021} & 88.73\mypm{0.09} & 81.81\mypm{0.29} & 95.39\mypm{0.05} & 90.63\mypm{1.39} & 90.34\mypm{1.14} \\
TransFuse~\cite{transfuse2021}   & 87.82\mypm{0.13} & 80.20\mypm{0.23} & 95.10\mypm{0.06} & 89.46\mypm{0.75} & 89.64\mypm{0.64} \\
H2Former~\cite{h2former2023}     & 88.61\mypm{0.22} & 81.77\mypm{0.31} & 95.52\mypm{0.14} & 91.36\mypm{0.73} & 89.45\mypm{0.33} \\
BRAU-Net++~\cite{lan2024brau}     & 89.59\mypm{0.72} & 83.04\mypm{1.37} & 95.68\mypm{0.09} & 91.46\mypm{0.40} & 91.02\mypm{1.72}
\\
\hline
SAM~\cite{sam2023}              & \cellcolor{lightgray}{49.75\mypm{0.00}} & \cellcolor{lightgray}{38.00\mypm{0.00}} & \cellcolor{lightgray}{60.38\mypm{0.00}} & \cellcolor{lightgray}{52.97\mypm{0.00}} & \cellcolor{lightgray}{79.79\mypm{0.00}} \\
MedSAM~\cite{Medsam2024}         & 85.09\mypm{0.54} & 76.03\mypm{0.68} & 93.82\mypm{0.39} & 87.12\mypm{0.58} & 87.71\mypm{0.52} \\
SAMed~\cite{SAMed2023}           & 89.75\mypm{0.45} & 83.12\mypm{0.56} & 95.81\mypm{0.23} & 91.61\mypm{0.24} & 90.60\mypm{0.69} \\
Med-SA~\cite{Wu2025Med-SA}       & 89.93\mypm{0.30} & 83.43\mypm{0.36} & 95.89\mypm{0.21} & 91.87\mypm{0.28} & \underline{91.35\mypm{0.27}} \\
SAMUS~\cite{samus2024}           & \underline{90.06\mypm{0.36}} & \underline{83.45\mypm{0.60}} & \underline{95.95\mypm{0.73}} & \underline{92.13\mypm{0.57}} & 90.83\mypm{0.37} \\
\hline
CVP-BAA                           & \textbf{90.40\mypm{0.20}} & \textbf{83.75\mypm{0.25}} & \textbf{96.27\mypm{0.36}} & \textbf{92.44\mypm{0.25}} & \textbf{91.66\mypm{0.46}} \\
\hline
\end{tabular}%
\end{table}

\subsubsection{Comparison on ISIC-2018}
\textbf{Quantitative Results}. Table~\ref{tab:isic} presents the quantitative evaluation on the ISIC-2018 dataset. Consistent with our observations on the CVC-ClinicDB dataset, CVP-BAA establishes a new state-of-the-art on this benchmark. Specifically, CVP-BAA significantly outperforms all competing CNN-based, Transformer-based, and existing SAM-adapted models, achieving superior performance with a Dice score of 90.40\%, IoU of 83.75\%, Accuracy of 96.27\%, Precision of 92.44\%, and Recall of 91.66\%. As a representative benchmark for dermoscopic imaging, a modality fundamentally distinct from breast ultrasound and endoscopy, these results further validate the robust cross-modality generalization capability of our proposed framework.

\begin{figure}[htbp]
    \footnotesize
    \imagestyle{0.96pt}{0.75}
     \newcommand{\imgw}{0.12\linewidth}

    \begin{tabular}{cccccccc}
        % row 1 
        \includegraphics[width=\imgw]{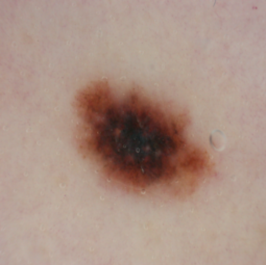} &
        \includegraphics[width=\imgw]{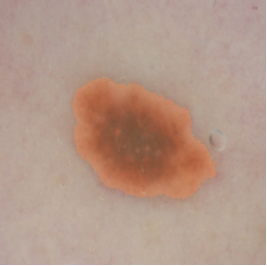} &
        \includegraphics[width=\imgw]{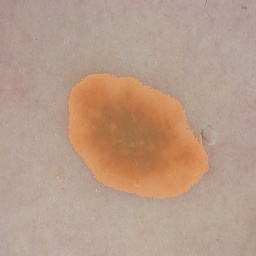} &
        \includegraphics[width=\imgw]{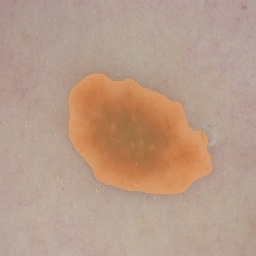} &
        \includegraphics[width=\imgw]{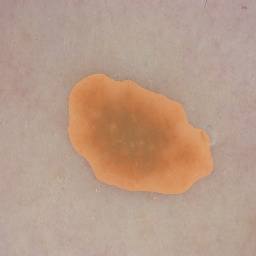} &
        \includegraphics[width=\imgw]{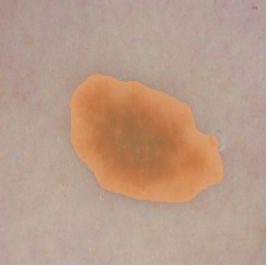} &
        \includegraphics[width=\imgw]{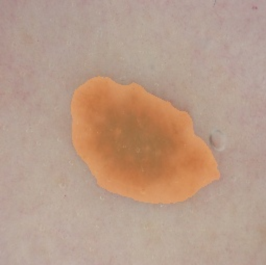} &
        \includegraphics[width=\imgw]{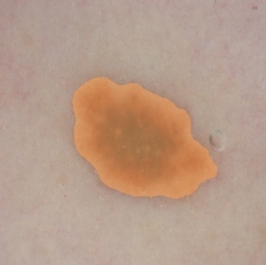} \\[-0.5mm]

        % row 2
        \includegraphics[width=\imgw]{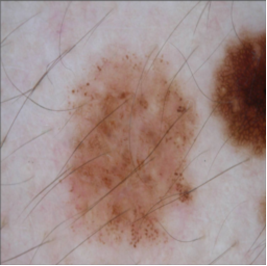} &
        \includegraphics[width=\imgw]{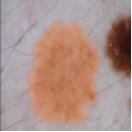} &
        \includegraphics[width=\imgw]{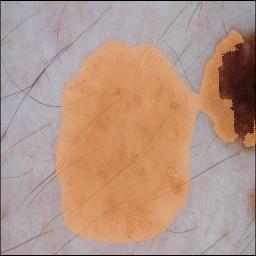} &
        \includegraphics[width=\imgw]{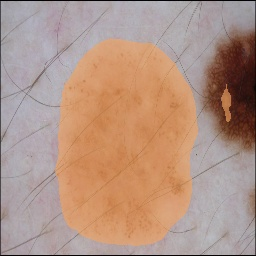} &
        \includegraphics[width=\imgw]{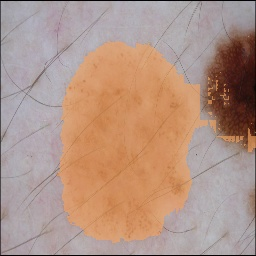} &
        \includegraphics[width=\imgw]{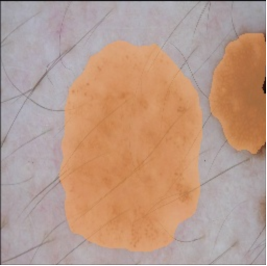} &
        \includegraphics[width=\imgw]{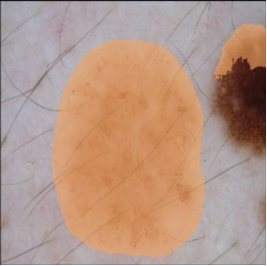} &
        \includegraphics[width=\imgw]{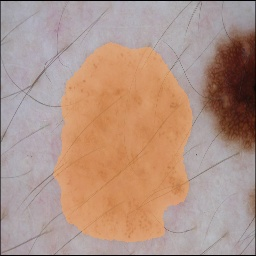}  \\
        
        Image &
        GT &
        UNet &
        AAU-Net &
        MISSFormer&
        H2Former &
        BRAU-Net++ &
        CVP-BAA \\
        \multicolumn{8}{c}{\textbf{(a)}}\\[1mm]

         \includegraphics[width=\imgw]{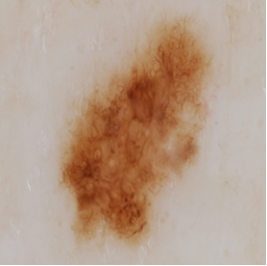} &
        \includegraphics[width=\imgw]{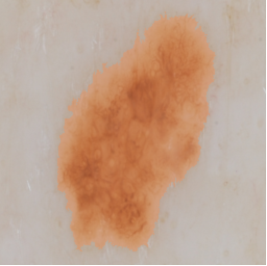} &
        \includegraphics[width=\imgw]{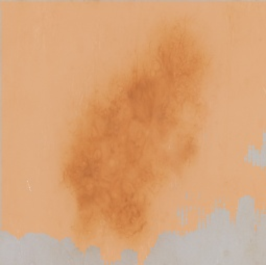} &
        \includegraphics[width=\imgw]{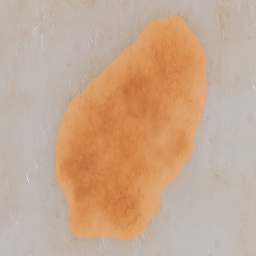} &
        \includegraphics[width=\imgw]{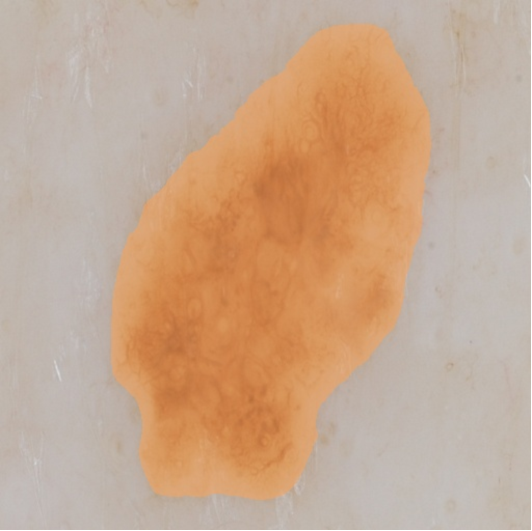} &
        \includegraphics[width=\imgw]{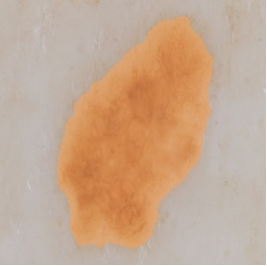} &
        \includegraphics[width=\imgw]{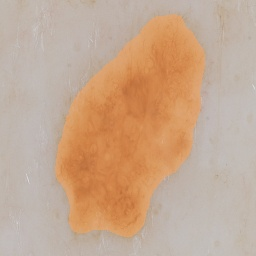} &
        \includegraphics[width=\imgw]{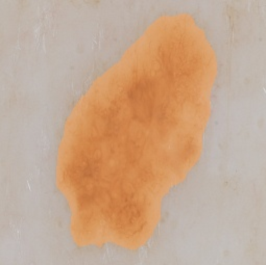} \\[-0.5mm]

        % row 2 
        \includegraphics[width=\imgw]{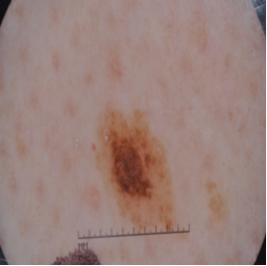} &
        \includegraphics[width=\imgw]{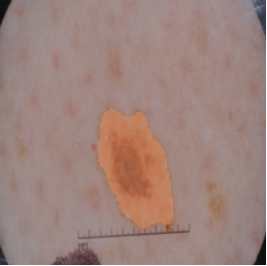}  &
        \includegraphics[width=\imgw]{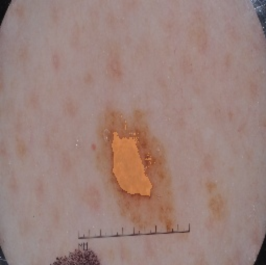} &
        \includegraphics[width=\imgw]{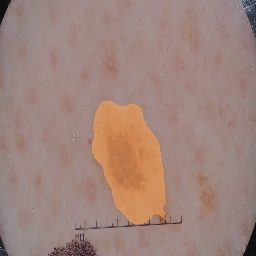} &
        \includegraphics[width=\imgw]{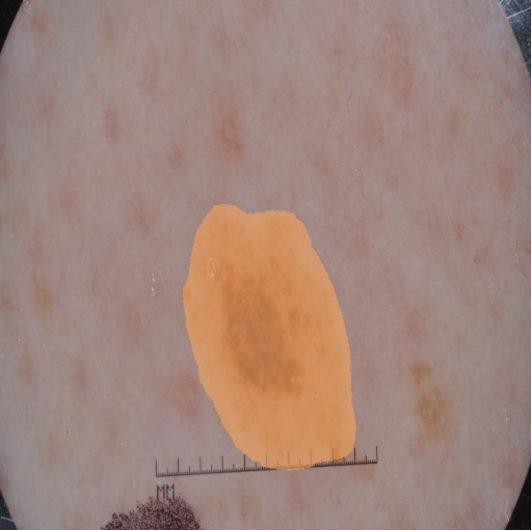} &
        \includegraphics[width=\imgw]{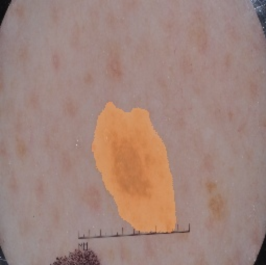} &
        \includegraphics[width=\imgw]{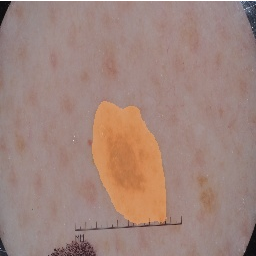} &
        \includegraphics[width=\imgw]{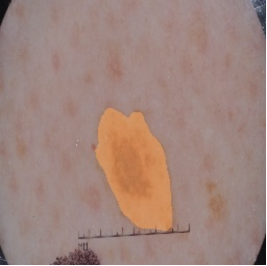} \\[-0.2mm]

        Image &
        GT  &
        SAM &
        MedSAM &
        SAMed &
        Med-SA &
        SAMUS &
        CVP-BAA \\
        \multicolumn{8}{c}{\textbf{(b)}}\\
    \end{tabular}
    \caption{Qualitative comparison of segmentation results on the ISIC-2018 dataset. The proposed CVP-BAA is compared against: (a) task-specific segmentation models, including UNet~\cite{unet2015}, AAU-Net~\cite{aaunet2022}, MISSFormer~\cite{missformer2021}, H2Former~\cite{h2former2023}, and BRAU-Net++~\cite{lan2024brau}; and (b) SAM and its adaptation methods, namely SAM~\cite{sam2023}, MedSAM~\cite{Medsam2024}, SAMed~\cite{SAMed2023}, Med-SA~\cite{Wu2025Med-SA}, and SAMUS~\cite{samus2024}. One can see that the delineation of both the lesion regions and their boundaries by CVP-BAA more closely aligns with the ground truth.
}
    \label{fig:isic_task_foundation}
\end{figure}

\textbf{Qualitative Results}. 
Visual comparisons in Fig.~\ref{fig:isic_task_foundation} strongly support these quantitative metrics. 
They demonstrate that our CVP and BAA modules effectively bridge domain gaps, leveraging CNN local inductive biases and transformer global modeling to ensure precise dermoscopic segmentation.

\subsection{Ablation Study}
\label{ablation_study}
In this section, we conduct a series of ablation studies on the BUSI and TN3K datasets to validate the effectiveness of individual components and their combinations. The quantitative results are presented in Table~\ref{tab:ablation}, and the visualizations are illustrated in Fig.~\ref{fig:gradcam_ablation_multi_dataset}. For our ablation study, we employ a modified version of SAMUS~\cite{samus2024} as the baseline. Specifically, this baseline omits the original cross-branch attention (CBA) and standard adapter modules, replacing the CBA mechanism with a simple element-wise summation to fuse CNN-derived features into the ViT encoder.

\subsubsection{Effectiveness of CVP}
\label{sec:ablation_cvp}
The integration of solely the CVP component into the baseline architecture leads to consistent and substantial segmentation gains: +2.12\% in Dice and -1.97 mm in HD on the BUSI dataset, and +2.30\% in Dice and -1.48 mm in HD on the TN3K dataset, respectively. These results validate that CVP effectively injects local fine-grained details to guide global dependency modeling and provides complementary cues for accurate lesion localization.

\subsubsection{Effectiveness of BAA}
\label{sec:ablation_baa}
Integrating the BAA component alone into the baseline yields consistent performance improvements: achieving a +3.58\% gain in Dice and a -2.83 mm reduction in HD on the BUSI dataset, alongside a +3.09\% gain in Dice and a -1.97 mm reduction in HD on the TN3K dataset. This indicates that incorporating the BAA significantly boosts performance, confirming its specific effectiveness in enhancing high-frequency boundary delineation and reducing boundary errors.

\subsubsection{Effectiveness of Combinations}
\label{sec:ablation_combination}
Combining both CVP and BAA achieves the best performance across both datasets, yielding 86.71\% Dice and 24.52 mm HD on BUSI, alongside 85.23\% Dice and 27.77 mm HD on TN3K. These results demonstrate that CVP and BAA work in concert, enabling a highly effective adaptation of SAM for medical image segmentation.

\begin{table}[htbp]
\centering
\footnotesize
\caption{Ablation study on the impact of each component and their combinations within CVP-BAA on the BUSI and TN3K datasets. The best result is highlighted in \textbf{bold}.}
\label{tab:ablation}
\begin{tabular}{@{} c c |cc|cc@{}}
\hline
\multicolumn{2}{c|}{Components} & \multicolumn{2}{c|}{BUSI} & \multicolumn{2}{c}{TN3K} \\
\cline{1-6}
 CVP& BAA & Dice (\%) $\uparrow$ & HD (mm) $\downarrow$ & Dice (\%) $\uparrow$ & HD (mm) $\downarrow$\\
\hline
$\xmark$ & $\xmark$  & 82.09\mypm{0.31} & 28.41\mypm{0.24} & 81.49\mypm{0.18} & 30.37\mypm{0.32} \\
$\cmark$& $\xmark$ & 84.21\mypm{0.35} & 26.44\mypm{0.31} & 83.79\mypm{0.31} & 28.89\mypm{0.25} \\
$\xmark$ &  $\cmark$ & 85.67\mypm{0.32} & 25.58\mypm{0.25} & 84.58\mypm{0.24} & 28.40\mypm{0.35} \\
$\cmark$ & $\cmark$  &  \textbf{86.71\mypm{0.33}} & \textbf{24.52\mypm{0.44}} & \textbf{85.23\mypm{0.30}} & \textbf{27.77\mypm{0.42}} \\
\hline
\end{tabular}
\end{table}

\subsubsection{Effectiveness of Prompt Injection Schemes}
To determine the optimal strategy for integrating convolutional visual prompts into the global multi-head self-attention (GMHSA) mechanism, we conduct an ablation study comparing two injection schemes: sequence prepending and element-wise addition. Specifically, we evaluate how different insertion methods into the key ($\bm{K}$) and value ($\bm{V}$) sequences of the ViT block impact overall segmentation performance. By default, as detailed in Section~\ref{subsec:GI_prompt_block}, we prepend the visual prompt $\bm{P}$ to the $\bm{K}$ and $\bm{V}$ sequences. An alternative approach is to add the prompts element-wise to $\bm{K}$ and $\bm{V}$, which preserves the original sequence length of the transformer. The quantitative results, presented in Table~\ref{tab:fusion_ablation}, demonstrate that our default sequence prepending strategy consistently outperforms the element-wise addition counterpart across both datasets, yielding gains of 0.78\% in Dice and -1.05 mm in HD on BUSI, and 0.49\% in Dice and -0.38 mm in HD on TN3K. Consequently, we adopt the sequence prepending scheme for our final architecture.

\begin{table}[htbp]
\centering
\footnotesize
\caption{Ablation study on the impact of convolutional visual prompt injection schemes. The best result is highlighted in \textbf{bold}.}
\label{tab:fusion_ablation}
\begin{tabular}{l|cc|cc}
\hline
\multirow{2}{*}{Schemes}  & \multicolumn{2}{c|}{BUSI} & \multicolumn{2}{c}{TN3K} \\
\cline{2-5}
& Dice (\%) $\uparrow$ & HD (mm) $\downarrow$& Dice (\%) $\uparrow$ & HD (mm) $\downarrow$\\
\hline
Add          & $85.93{\pm}0.75$ & $25.57{\pm}0.53$ & $84.74{\pm}0.28$ & $28.15{\pm}0.26$ \\
Prepend       & $\mathbf{86.71{\pm}0.33}$ & $\mathbf{24.52{\pm}0.44}$ & $\mathbf{85.23{\pm}0.30}$ & $\mathbf{27.77{\pm}0.42}$ \\
\hline
\end{tabular}
\end{table}

\subsubsection{Effectiveness of Prompt Injection Positions }
We conduct ablation experiments to investigate the impact of injecting the CVP strategy at different depths of the ViT encoder, specifically at the \textit{3}-rd, \textit{6}-th, \textit{9}-th, and \textit{12}-th global ViT blocks. The ablation scheme and corresponding results are summarized in Table~\ref{tab:cvp_position}. Empirically, expanding the injection to more blocks enables the frozen SAM encoder to continuously integrate fine-grained local priors from the CNN branch throughout its depth. This progressive feature fusion culminates in the full injection configuration, i.e., all the \textit{3}-rd, \textit{6}-th, \textit{9}-th, and \textit{12}-th blocks, which consistently obtains the highest Dice scores and the lowest HD values on both BUSI and TN3K datasets.

\begin{table}[htbp]
\centering
  \footnotesize
  \caption{Ablation study on the impact of  convolutional visual prompt injection positions. The best results are marked in \textbf{bold}.}
  \label{tab:cvp_position}
  \begin{tabular}{l|cc|cc}
    \hline
    \multirow{2}{*}{Configurations}  & \multicolumn{2}{c|}{BUSI} & \multicolumn{2}{c}{TN3K} \\
    \cline{2-5}
    & Dice (\%) $\uparrow$ & HD (mm) $\downarrow$ & Dice (\%) $\uparrow$ & HD (mm) $\downarrow$\\
    \hline
     \textit{3}-rd & $86.23{\pm}0.53$ & $25.83{\pm}1.32$ & $84.74{\pm}0.27$ & $28.58{\pm}0.43$ \\
     \textit{3}-rd, \textit{6}-th &  $86.13{\pm}0.49$ & $25.37{\pm}0.43$  & $84.71{\pm}0.29$ & $28.15{\pm}0.36$ \\
     \textit{3}-rd, \textit{6}-th, \textit{9}-th & $86.26{\pm}0.32$ & $25.13{\pm}0.38$  & $84.97{\pm}0.35$ & $28.01{\pm}0.40$   \\
     \textit{3}-rd, \textit{6}-th, \textit{9}-th, \textit{12}-th& $\mathbf{86.71{\pm}0.30}$ & $\mathbf{24.52{\pm}0.44}$ & $\mathbf{85.23{\pm}0.30}$ & $\mathbf{27.77{\pm}0.42}$ \\
    \hline
  \end{tabular}
\end{table}

\subsubsection{Effectiveness of Scaling Factor s}
We conduct ablation studies to evaluate the sensitivity of our method to the scaling factor $s \in \{0.1, 0.3, 0.5, 0.7, 1.0\}$. As shown in Table~\ref{tab:ablation_scaling}, the model peaks at $s=0.3$, achieving the best Dice and HD metrics across both datasets. The performance trend reveals a clear trade-off: smaller scaling factors inject insufficient task-specific information, limiting the adapter's effectiveness, while larger scaling factors aggressively perturb the pretrained backbone, leading to optimization instability and degraded generalizability. 
Therefore, an appropriate scaling factor acts as a crucial regularizer that balances the contribution of the high-frequency adapter branch against the primary transformer pathway. Consequently, we fix $s=0.3$ for all our experiments.

\begin{table}[htbp]
\centering
\footnotesize
\caption{Ablation study on the boundary enhancement (BE) module and the branch scaling factor $s$. The best results are highlighted in \textbf{bold}.}
\label{tab:ablation_scaling}
\begin{tabular}{l|c|cc|cc}
\hline
\multirow{2}{*}{BE} & \multirow{2}{*}{$s$} & \multicolumn{2}{c|}{BUSI} & \multicolumn{2}{c}{TN3K} \\
\cline{3-6}
& & Dice (\%) $\uparrow$ & HD (mm) $\downarrow$ & Dice (\%) $\uparrow$ & HD (mm) $\downarrow$\\
\hline
\multirow{5}{*}{w/}
& 0.1 & 85.39\mypm{0.77} & 26.36\mypm{0.82} & 84.90\mypm{0.30} & 27.94\mypm{0.42} \\
& 0.3 & \textbf{86.71\mypm{0.33}} & \textbf{24.52\mypm{0.44}} & \textbf{85.23\mypm{0.30}} & \textbf{27.77\mypm{0.42}} \\
& 0.5 & 85.04\mypm{0.20} & 25.75\mypm{0.81} & 84.91\mypm{0.18} & 27.96\mypm{0.56} \\
& 0.7 & 85.44\mypm{0.51} & 25.84\mypm{0.29} & 84.76\mypm{0.41} & 27.93\mypm{0.60} \\
& 1.0 & 84.55\mypm{1.18} & 27.64\mypm{1.90} & 84.78\mypm{0.15} & 27.79\mypm{0.17} \\
\hline
\multirow{5}{*}{w/o}
& 0.1 & 82.16\mypm{1.35} & 28.52\mypm{1.92} & 80.90\mypm{0.55} & 32.05\mypm{0.71} \\
& 0.3 & 81.83\mypm{0.88} & 28.99\mypm{1.12} & 80.57\mypm{0.40} & 31.89\mypm{0.39} \\
& 0.5 & 81.47\mypm{0.75} & 29.61\mypm{1.20} & 80.40\mypm{0.36} & 32.61\mypm{0.72} \\
& 0.7 & 82.19\mypm{0.39} & 28.76\mypm{0.35} & 80.35\mypm{0.23}&32.50\mypm{0.52} 
\\
& 1.0 & 80.94\mypm{0.10} & 31.00\mypm{1.94} & 80.17\mypm{0.56} & 32.74\mypm{0.36}
\\
\hline
\end{tabular}
\end{table}

\begin{figure}[htbp]
    \centering
    \tablestyle{0.96pt}{0.85}
     \newcommand{\imgw}{0.16\linewidth}

    \begin{tabular}{ccccccc}
      \footnotesize
         & Image & GT & (a) Baseline & (b) w/ CVP & (c) w/ BAA & (d) CVP-BAA  \\
       
        % \multirow{2}{*}{BUSI}
        \multirow{2}{*}{\rotatebox[origin=c]{90}{BUSI}}
        & \includegraphics[width=\imgw]{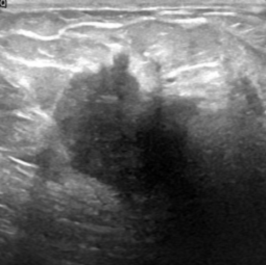}
        & \includegraphics[width=\imgw]{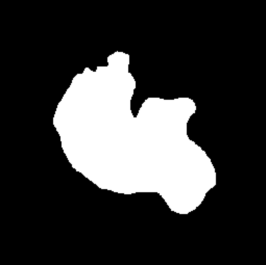} 
        & \includegraphics[width=\imgw]{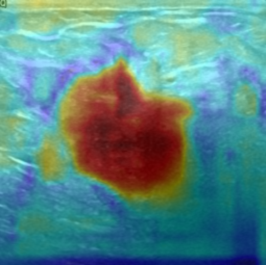}
        & \includegraphics[width=\imgw]{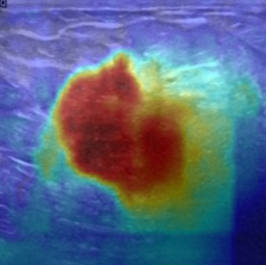}
        & \includegraphics[width=\imgw]{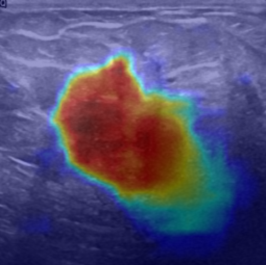}
        & \includegraphics[width=\imgw]{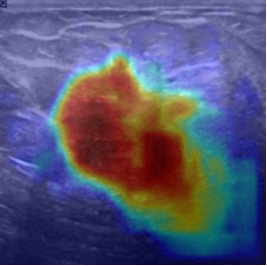}\\[-0.5mm]

        & \includegraphics[width=\imgw]{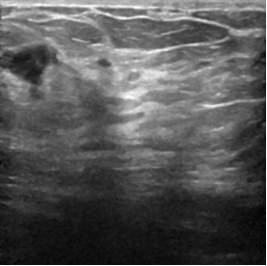}
        & \includegraphics[width=\imgw]{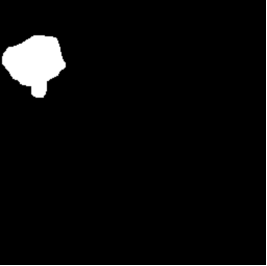}
        & \includegraphics[width=\imgw]{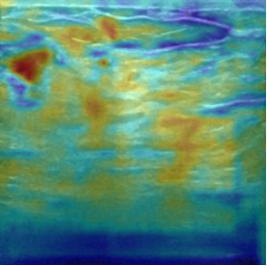}
        & \includegraphics[width=\imgw]{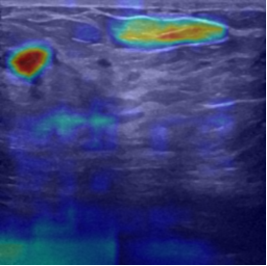}
        & \includegraphics[width=\imgw]{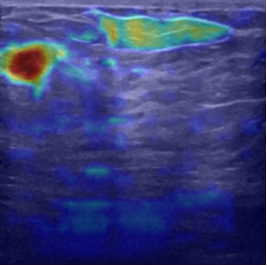}
        & \includegraphics[width=\imgw]{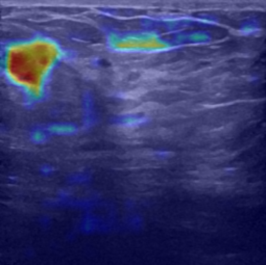} \\[0.5mm]

        % \multirow{2}{*}{TN3K}
        \multirow{2}{*}{\rotatebox[origin=c]{90}{TN3K}}
        & \includegraphics[width=\imgw]{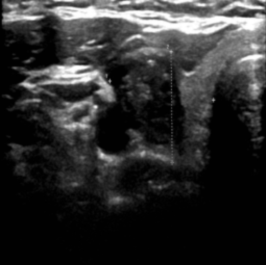}
        & \includegraphics[width=\imgw]{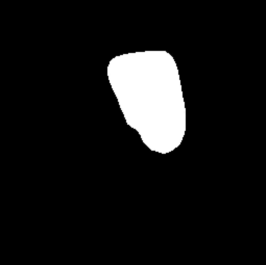}
        & \includegraphics[width=\imgw]{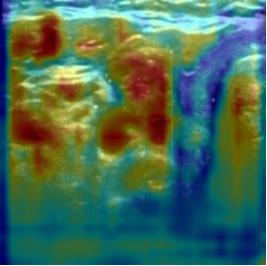}
        & \includegraphics[width=\imgw]{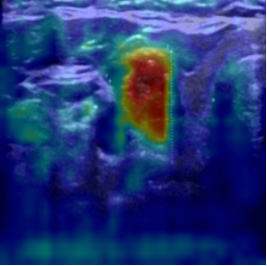}
        & \includegraphics[width=\imgw]{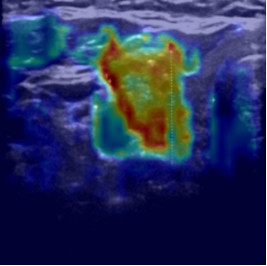}
        & \includegraphics[width=\imgw]{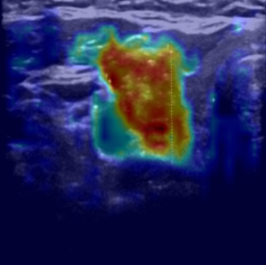} \\[-0.5mm]

        & \includegraphics[width=\imgw]{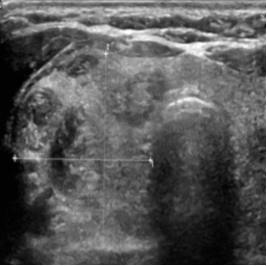}
        & \includegraphics[width=\imgw]{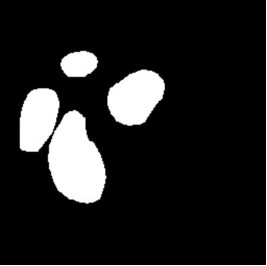} 
        & \includegraphics[width=\imgw]{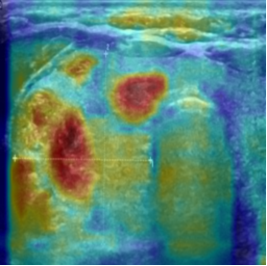}
        & \includegraphics[width=\imgw]{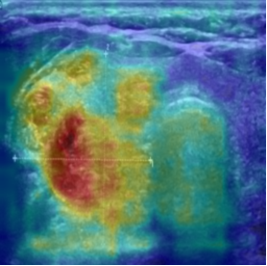}
        & \includegraphics[width=\imgw]{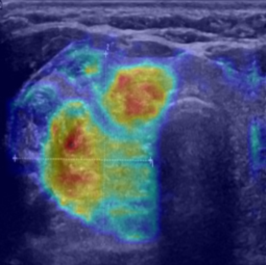}
        & \includegraphics[width=\imgw]{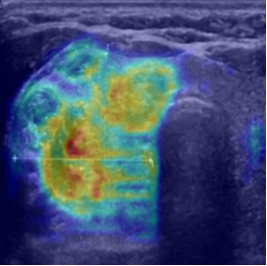}\\
    \end{tabular}
     \caption{Grad-CAM visualizations illustrating the attention maps of different components and their combinations on the BUSI and TN3K datasets. (a) Baseline (w/o CVP \& BAA) shows dispersed attention focusing mainly on the low-frequency central regions, missing ambiguous boundaries. (b) With CVP, the model learns to attend to local fine-grained structural priors. (c) With BAA, the high-frequency boundary regions are significantly activated, confirming the effectiveness  of our explicit high-pass filtering design.
     (d) The full CVP-BAA model yields more precise and boundary-aligned attention maps, visually corroborating the findings observed in our quantitative ablation studies in Sections~\ref{sec:ablation_cvp},~\ref{sec:ablation_baa}, and~\ref{sec:ablation_combination}.
     }
\label{fig:gradcam_ablation_multi_dataset}
\end{figure}

\subsubsection{Visualization of Component Contributions}
To provide qualitative insights into the internal representations learned by each proposed component, we visualize the attention maps from the final block of the SAM image encoder using Grad-CAM~\cite{selvaraju2017grad}, as illustrated in Fig.~\ref{fig:gradcam_ablation_multi_dataset}. The attention distribution of the baseline model is highly dispersed and frequently focuses on irrelevant background regions, leading to inaccurate lesion localization. In contrast, incorporating the CVP enhances attention on local, fine-grained structures, such as the spiculated margins of malignant breast tumors in BUSI or the subtle textural variations of thyroid nodules in TN3K. This demonstrates that our local CNN features successfully guide the global attention mechanism. Meanwhile, the integration of the BAA explicitly activates high-frequency boundary regions, directly validating our design motivation of isolating edge cues via high-pass filtering. Finally, the full CVP-BAA model integrates these complementary strengths, refining the overall attention distribution to produce the most precise and boundary-aligned activation maps, thereby providing compelling visual evidence of our framework's effectiveness.

\subsection{Efficiency Analysis}
We further evaluate the computational efficiency of the compared SAM-based models in terms of trainable parameters (\#Train. Params), total parameters (\#Total Params), computational complexity (GFLOPs), and inference speed (FPS). To ensure a fair comparison, all metrics are measured on the BUSI dataset using a uniform input resolution of 256$\times$256 under identical hardware configurations, as summarized in Table~\ref{tab:efficiency_comparison}. Compared to the strongest baseline, SAMUS, CVP-BAA significantly improves parameter efficiency by reducing the trainable parameter count from 39.65M to 30.20M, a $\sim$23.8\% reduction, while maintaining a smaller total parameter scale. Furthermore, CVP-BAA achieves a highly competitive computational footprint, requiring only 143.46 GFLOPs and delivering a robust inference speed of 37.92 FPS. Although requiring more trainable parameters (with 30.20M) than SAMed (with 18.81M), CVP-BAA achieves a superior trade-off between parameter efficiency and model capacity, maintaining comparable inference speed while yielding substantially better segmentation accuracy, as shown in Table~\ref{tab:compression_busi_tn3k_udiat}. These results demonstrate that CVP-BAA enhances adaptation effectiveness without introducing excessive computational overhead.

\begin{table}[htbp]
\centering
\footnotesize
\caption{Computational efficiency comparison among the evaluated SAM‑based models. The symbol $\uparrow$ indicates the larger the better. The symbol $\downarrow$ denotes the smaller the better.}
\label{tab:efficiency_comparison}
\begin{tabular}{l|c|c|c|c}
\hline
Methods & Train. Params (M) $\downarrow$ & Total Params (M) $\downarrow$ & GFLOPs $\downarrow$ & FPS $\uparrow$ \\
\hline
SAM~\cite{sam2023}        & 0.00  & 90.49  & 371.99 & 12.10 \\
MedSAM~\cite{Medsam2024}  & 4.06  & 90.49  & 371.99 & 12.10 \\
SAMed~\cite{SAMed2023}   & 18.81 & 90.36  & 103.41 & 37.06 \\
Med‑SA~\cite{Wu2025Med-SA}& 13.00 & 97.59  & 402.88 & 11.55 \\
SAMUS~\cite{samus2024}   & 39.65 & 130.10 & 145.88 & 41.78 \\
CVP‑BAA                  & 30.20 & 127.73 & 143.46 & 37.92 \\
\hline
\end{tabular}
\end{table}

\section{Conclusion}
\label{sec:conclusion}
In this paper, we propose CVP-BAA, an efficient SAM adaptation framework for medical image segmentation. The proposed convolutional visual prompting (CVP) strategy explicitly injects local structural priors into the global dependency modeling. Furthermore, the boundary-aware adapter (BAA) significantly enhances the representation of fine-grained structures and ambiguous lesion boundaries. Extensive experiments on multiple public datasets across diverse imaging modalities demonstrate that CVP-BAA consistently outperforms both dedicated task-specific segmentation models and existing SAM-based adaptation methods. These results validate the effectiveness of combining explicit local priors with boundary-aware structural refinement, highlighting the potential of foundation model adaptation in medical image analysis. Nevertheless, the current design introduces additional architectural components, such as the BAA, which inevitably incurs a marginal increase in learnable parameters. Furthermore, the model's generalization capabilities in completely unseen clinical domains and data-scarce scenarios remain open challenges. Future work will focus on developing even more parameter-efficient mechanisms and exploring few-shot learning paradigms to enhance robustness in low-resource medical environments.

\section*{Data Availability Statement}
All data that support the findings of this study are included within the article.

\section*{CRediT Authorship Contribution Statement}
\textbf{Mei Wang}: Data Curation, Investigation, Methodology, Software, Visualization, Writing – Original Draft Preparation, Writing – Review \& Editing. \textbf{Minghui Li}: Investigation, Validation, Visualization, Writing – review \& editing. \textbf{Libin Lan}: Conceptualization, Funding Acquisition, Methodology, Project Administration, Software, Writing – Original Draft Preparation, Writing – Review \& Editing.

\section*{Declaration of Competing Interest}
The authors declare that they have no known competing financial interests or personal relationships that could have appeared to influence the work reported in this paper.

\section*{Acknowledgments} 
This work described in this paper was supported in part by the Scientific Research Foundation of Chongqing University of Technology under Grant 2021ZDZ030 and in part by the Youth Project of Science and Technology Research Program of Chongqing Education Commission of China under Grant KJQN202301145.

{
\balance
\bibliographystyle{elsarticle-num}
\bibliography{refs}
}

\end{document}